\documentclass[twocolumn]{article}

\usepackage{graphicx}
\usepackage{amsmath,amssymb}
\usepackage{booktabs}
\usepackage{tabularx}
\usepackage{longtable}
\usepackage{float}
\usepackage{enumitem}
\usepackage{xcolor}
\usepackage{algorithm}
\usepackage{algpseudocode}
\usepackage{url}
\usepackage[neutral]{naturestyle}%
\usepackage{placeins}
\makeatletter
\let\figure\nat@figure
\let\endfigure\nat@endfigure
\makeatother
\usepackage{etoolbox}
\newtoggle{comment}
\togglefalse{comment}

\newcommand{\method}{s$^3$}

\makeatletter
\newcommand\nat@teaser{}
\newcommand\teaserfigure[1]{\renewcommand\nat@teaser{%
  \vskip 14pt\begin{minipage}{\textwidth}\centering #1\end{minipage}}}
\patchcmd{\nat@titlepage}{\vskip 26pt}{\nat@teaser\vskip 26pt}{}{\PackageWarning{workshop}{teaser patch failed}}
\makeatother

\graphicspath{{workshop_figures/}{toy_results/}}
\newcommand{\sweepfigure}[2]{%
  \IfFileExists{workshop_figures/#1}{\includegraphics[width=#2]{#1}}{%
    \fbox{\parbox[c][0.28\textheight][c]{#2}{\centering\small Placeholder: \texttt{workshop\_figures/#1}}}}}

\begin{document}
\addtocontents{toc}{\protect\setcounter{tocdepth}{-10}}

\title{s$^3$: Simple Scaling of Swarms via Cooperative Multi-agent RL}
\runningfooter{s$^3$: Simple Scaling of Swarms}

\author{
  Kunal Jha\textsuperscript{1,\textdagger},
  Gon\c{c}alo Faria\textsuperscript{1,2,\textdagger},
  Shuyue Stella Li\textsuperscript{1,\textdagger},
  Scott Geng\textsuperscript{1,\textdagger},
  Faeze Brahman\textsuperscript{2},
  Noah A. Smith\textsuperscript{1,2},
  Yulia Tsvetkov\textsuperscript{1},\\
  Max Kleiman-Weiner\textsuperscript{1,\textdaggerdbl} \&
  Natasha Jaques\textsuperscript{1,\textdaggerdbl}
}
\affiliations{\textsuperscript{1}University of Washington. \textsuperscript{2}Allen Institute for Artificial Intelligence.}
\authornotes{\textsuperscript{\textdagger}Equal contribution. \textsuperscript{\textdaggerdbl}Equal advising. Correspondence to \href{mailto:kjha@uw.edu}{kjha@uw.edu}.}

\maketitle
\raggedbottom

\teaserfigure{
  \vspace{-2mm}
  \includegraphics[width=0.75\textwidth]{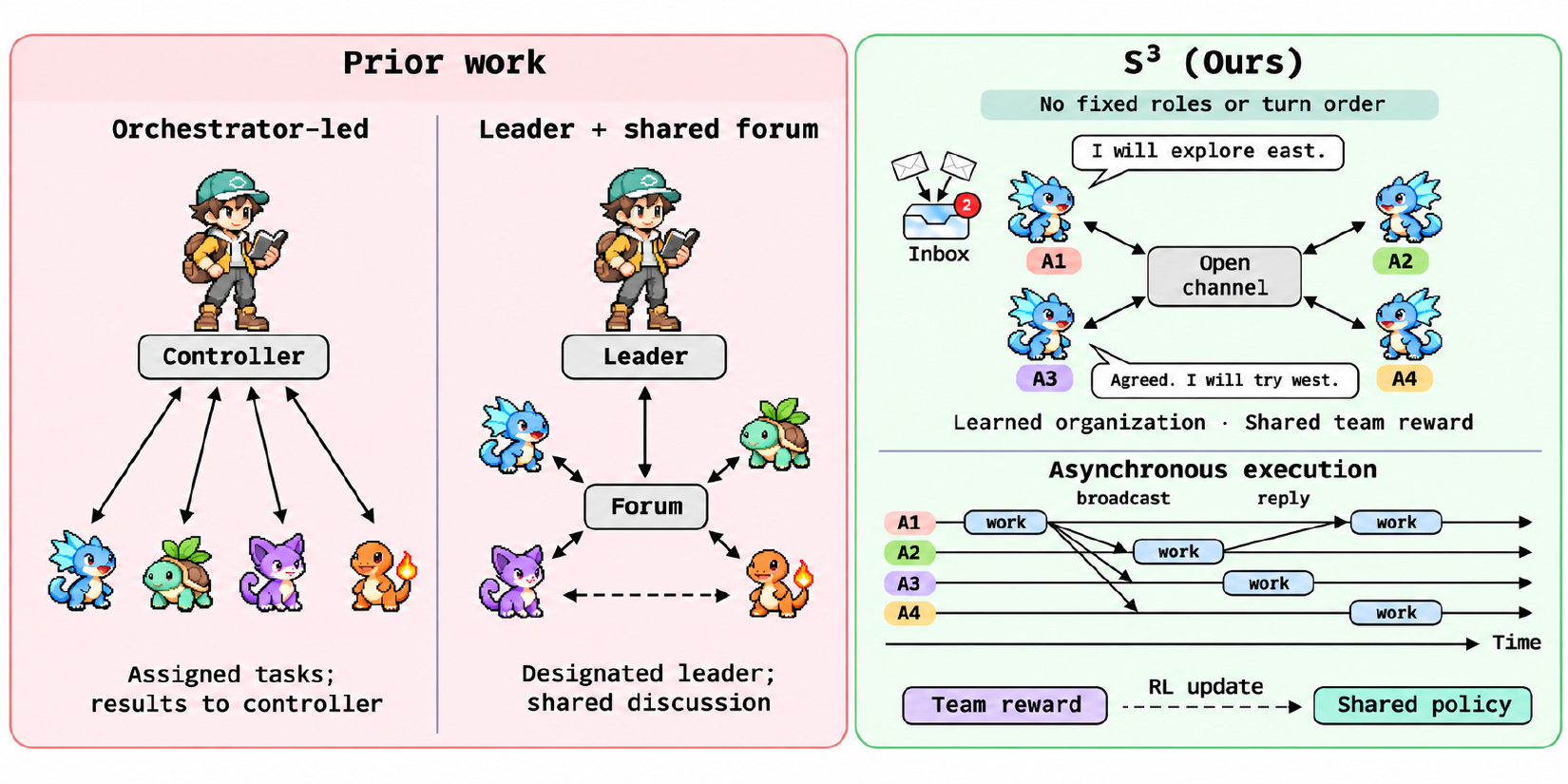}\vspace{-2mm}
  \captionof{figure}{\textbf{Left:} prior multi-agent systems organize agents with static harnesses designed in advance, such as a controller that assigns tasks or a leader that directs a shared forum. \textbf{Right:} \method{} is the first multi-agent reinforcement learning method in which agents decide for themselves how long to work alone and when to share findings or read teammates’ messages, with no predefined organization or roles. Agents are trained with a shared team reward, and their organization emerges from training.}
  \label{fig:teaser}}

\begin{abstract}

Groups can achieve more than individuals in the same amount of time, but only if their members coordinate: they must decide who works on what and what to share. However, this organization in multi-agent systems today usually comes from arbitrary, static harnesses designed a priori without adapting to agent behaviors or task specifications. We ask whether agents can instead learn to organize themselves.
We introduce \method{}, the first multi-agent reinforcement learning method in which agents decide for themselves how long to work alone and when to share findings or read teammates’ messages, with no predefined organization or roles.
Our \method{} agents learn to take different approaches, read each other's work, exhibit emergent roles that complement each other in the swarm, and generalize to bigger teams.
We study both coordination, where agents must combine private information, and parallel exploration, where sharing discoveries can make search more efficient.
With matched training and inference budgets, \method{} swarms evaluated with twice as many agents as in training solve over 25$\times$ more coordination problems than the strongest baseline, and in exploration, communicating agents reduce repeated work by nearly 50\% and outperform best-of-$N$ by up to 9.4 percentage points.
Rather than being designed in advance, \method{} enables organization to emerge from a shared goal. This opens the door to capable systems built from many models that learn to work together.
\end{abstract}

\section{Introduction}

Intelligence has always been a collective achievement: human brains are thought to have evolved for life in groups, and human knowledge has grown by being shared and built upon across generations \citep{humphrey1976social,dunbar1998social,henrich2016secret}. Artificial intelligence has recently begun to scale the same way, by running many language models (LMs) at once. Tens of thousands of agents were deployed to prove a long outstanding theorem about the Navier--Stokes problem \citep{openaiNavierStokesMillennium}, agent swarms split a task across parallel subagents to cut latency \citep{kimiteam2026kimik25visualagentic}, and agents that share findings at test time discover solutions that independent attempts miss \citep{park2026scalingdiscoverytesttimecommunication}. Many multi-agent systems use an organization specified by their designers. A central orchestrator decomposes the task and merges results \citep{wu2023autogenenablingnextgenllm,fourney2024magenticonegeneralistmultiagentsolving,anthropicBuiltMultiagent}, agents debate in rounds \citep{du2023improvingfactualityreasoninglanguage}, or each agent is given a role \citep{hong2024metagptmetaprogrammingmultiagent}.

Hand-designed organizations are fixed guesses about how a team should work, and they often break down. Orchestrated systems duplicate work, flood each other with updates, and fail to reconcile subagents that disagree \citep{anthropicBuiltMultiagent,cemri2025multiagentllmsystemsfail}. Human organization, by contrast, is adaptive. How people divide labor, specialize, and share information depends on the task, and much of this structure emerged through cultural evolution rather than by design \citep{henrich2016secret,Boyd1985-gr}. Centrally planned economies often struggled in part because no planner could gather what the people doing the work knew \citep{hayek}. A fixed harness, or even a prompted one, can likewise miss the specializations a task calls for. The bitter lesson of AI is that learned behavior repeatedly overtakes behavior designed by hand \citep{incompleteideasBitterLesson,Silver2021-af}: reasoning models, for example, improved most by learning to reason through reinforcement learning on checkable outcomes, rather than by following reasoning strategies written by people \citep{openai2026openaio1card,Guo_2025}. We think the same holds for multi-agent organization. Rather than imposing our own ideas of how agents should interact, we let the organization emerge from training. If a central coordinator is the best way to solve a task, the team should learn to create one; if a task rewards specialization, agents should learn to divide the work. Because each agent works on its own copy of the task, agents can coordinate only through what they tell each other, so learning to organize means learning to communicate: what to share, with whom, and when sharing is not worth its cost.

To enable agents to learn to organize themselves, we introduce \method{}, a method for training and scaling swarms of language model agents (Figure~\ref{fig:teaser}). A team is $N$ copies of one model. No agent is a leader, no agent has a role, and there is no turn order. All agents work at the same time, each on its own copy of the task, and whenever it chooses, an agent can message any teammate or all of them and read the messages others have sent it. Any organization can arise from this setting, including a single leader that collects results and directs the others. \method{} trains the shared policy with group-relative reinforcement learning, with every agent receiving the same team reward. For each problem, we sample independent team attempts, normalize their team rewards, and assign every token an agent generates its team's advantage. This trains task actions and communication together, so an agent can be rewarded for helping a teammate succeed. For training to reinforce effective communication, agents must be able to act on one another's discoveries while they are still useful. Our infrastructure interleaves agents' computation, handling tool calls and message delivery asynchronously so that a slow agent does not hold up its partners (Section~\ref{sec:setting}).

A trained swarm is worth having only if it beats $N$ agents working alone. This best-of-$N$ baseline is strong, it allows agents to use additional compute and empirically, best-of-$N$ performance improves as $N$ grows \citep{brown2024largelanguagemonkeysscaling,wang2023selfconsistencyimproveschainthought,snell2024scalingllmtesttimecompute}. Thus, a swarm can appear to perform better simply because it runs more agents, even if its members never help each other, and these gains can shrink or disappear when budgets are matched \citep{generalreasoning2026avalonm1,jha2026solosociallearningcharacterizing}. With learning and inference compute matched to single-agent counterparts, we ask: \emph{how can we train swarms to become more capable as they grow by learning to communicate, coordinate, and build on one another's discoveries?}

We study two tasks that test how swarms share and use information. First, we explore an information aggregation task, which tests whether agents can coordinate to combine private information held by different members of the swarm. We then study decentralized exploration, where agents must discover information through interaction and can share what they learn to divide the search and avoid repeating work. This models collaborative discovery, where findings from one search can guide another \citep{Stanley2015-nl}. Across both tasks, \method{} swarms coordinate where imitation and individual rewards fail, keep working with more agents than they were trained with, and divide their search to outperform best-of-$N$. Our contributions and findings are:

\begin{enumerate}[leftmargin=*,topsep=2pt,itemsep=1pt,parsep=0pt,partopsep=0pt]
\item \textbf{Learning coordination through cooperative MARL:} to our knowledge, we provide the first demonstration that cooperative multi-agent reinforcement learning (MARL) can train asynchronous LM swarms whose emergent coordination outperforms best-of-$N$ sampling from solo policies trained with matched rollout budgets and per-agent token allowances, without prescribed roles or communication structures (Section~\ref{sec:results}).
\item \textbf{Careful benchmarking of learned collaboration:} by comparing against independent agents with matched training and token budgets, and by ablating communication, the shared reward, and demonstrations, we show that the gains come from learning to coordinate (Section~\ref{sec:results}).
\item \textbf{Generalization to larger numbers of agents:} swarms retain an advantage over the controls when evaluated with more agents than they were trained with, showing how learned coordination can support scaling at inference without retraining (Section~\ref{sec:results}).
\item \textbf{An account of how organization emerges:} training trajectories reveal how shared rewards favor efficient information exchange and coordinated exploration, while individual rewards create a social dilemma that undermines cooperation (Section~\ref{sec:analysis}).
\end{enumerate}

\section{Related work}
\label{sec:related-main}
 Prompted multi-agent systems benefit from specialized roles, discussion, and coordinated workflows, but their interaction patterns are specified by their designers \citep{li2023camelcommunicativeagentsmind,qian2024chatdevcommunicativeagentssoftware,hong2024metagptmetaprogrammingmultiagent,chen2023agentversefacilitatingmultiagentcollaboration,wang2024mixtureofagentsenhanceslargelanguage}. More recent work searches over or evolves agent workflows and harnesses to improve performance while keeping model weights fixed \citep{zhang2025aflowautomatingagenticworkflow,hu2025automateddesignagenticsystems,zhang2025multiagentarchitecturesearchagentic,zhang2025gdesignerarchitectingmultiagentcommunication,zhuge2024languageagentsoptimizablegraphs}. Others reorganize the team itself, assigning agents with a learned reward \citep{zhou2025resorewarddrivenselforganizingllmbased}, restructuring the team through a central supervisor \citep{cao2026evolvingagentteams}, or learning reusable teamwork strategies without a fixed workflow \citep{pappu2026selforganizingagentteamslearn}. Multi-agent RL has also trained collaboration within debate, assigned roles, synchronous interactions, or team decisions with information split across agents \citep{park2025maporlmultiagentpostcotrainingcollaborative,motwani2025maltimprovingreasoningmultiagent,liu2025llmcollaborationmultiagentreinforcement,wan2025remalearningmetathinkllms,park2026teamgymgymteamdecisionmaking}. Closest to our setting, Avalon-M1 jointly trains agents with a collective reward, but builds in a CEO that hires specialists and organizes their work \citep{generalreasoning2026avalonm1}; it does not establish a coordination gain over best-of-$N$ sampling from a solo policy trained with a matched budget. To our knowledge, we are the first to demonstrate this gain in asynchronous swarms without prescribed roles or communication structures. Appendix~\ref{sec:related} provides the full related-work discussion.

\begin{table}[t]
{\sffamily\centering
\small
\caption{Controls and baselines. Swarm controls are ablations of \method{}; individual-reward swarms still communicate and learn together, whereas solo agents train independently. SFT denotes supervised fine-tuning. The SFT ablation applies to decentralized exploration; information aggregation starts from the base model. Team learners train with four agents and solo learners with one. Within each task, matched learners use the same problems, number of agent trajectories per update, per-agent token allowance, and optimizer. Rewards include the token cost in Equation~\ref{eq:reward}.}
\label{tab:conditions}
\begin{tabularx}{\linewidth}{@{}lX@{}}
\toprule
Condition & Description \\
\midrule
Solo (SFT+RL) & Train a single agent with individual reward; evaluate independent copies as best-of-$N$ \\
\method{}: no communication & Team reward with messaging tools disabled \\
\method{}: individual reward & Messaging enabled; each agent is rewarded for its own success \\
\method{}: online SFT & Train on successful communicating teams sampled from the current policy \\
SFT only / initial & Fine-tuned checkpoint or base model, without RL \\
\method{}: no SFT & Team reward and messaging, trained from the base model \\
Solo RL (no SFT) & Single-agent RL from the base model, evaluated as best-of-$N$ \\
\bottomrule
\end{tabularx}\par}
\end{table}

\section{Methods}
\label{sec:setup}
\label{sec:method}

\subsection{Setting and harness}
\label{sec:setting}

A problem $x$ is attempted by a team of $N$ agents. All agents share one policy $\pi_\theta$, but each agent $i$ has its own history, its own private observations, and its own copy of the task tools. Agent $i$ has a limit of $L$ generated tokens in total, counting reasoning, tool calls, and messages. The whole team receives one reward.

To communicate, each agent has a tool called \texttt{message(to, text)}, which sends text to one teammate or to all, and a tool called \texttt{read\_messages()}, which returns its unread messages. Agents choose whether to read what teammates wrote. There is no limit on the number of tool calls or messages within the token allowance, and no separate reward or penalty for messaging. Teammates are addressed by public names sampled without replacement from \texttt{agent\_1} to \texttt{agent\_64} in each episode, and each agent sees its teammates in a shuffled order, so cannot learn conventions dependent on a particular teammate name. The full tool definitions appear in Appendix~\ref{app:harness}.

Agents use these tools asynchronously: each generates text, calls tools, and resumes when its own tool result returns. One agent can therefore keep exploring or share a discovery while another is waiting for a tool, instead of the entire swarm progressing at the pace of its slowest member. When a message arrives, the harness inserts a notification into the recipient's history as a tool output, listing unread counts by sender. If the recipient is generating text, the notification is inserted after its current inference request finishes; an idle recipient is woken immediately. The notification tells the agent that a message is waiting, and \texttt{read\_messages()} supplies its contents. An agent that ends a response without a tool call becomes idle and spends no tokens until another message arrives (Appendix~\ref{app:harness}).

As the swarm grows, its agents compete for the same inference server. The server has a fixed number of slots for concurrent sequences, a token budget per inference batch, and limited KV-cache memory for retaining each agent's history. Long generations occupy sequence slots, while long histories consume KV-cache capacity, limiting how many agents can generate at once. If a few agents monopolize these resources, their teammates wait, and an agent may finish most of its search before its partners can contribute. We therefore split generation into requests of at most 1,024 new tokens. A request ends early at a tool call or the end of a response; otherwise the agent continues from its exact generated prefix. These boundaries give other agents opportunities to progress and allow pending notifications to be inserted, without adding a new turn or an instruction to the model. Agents can thus respond to one another during their work without a fixed turn order (Appendices~\ref{app:harness} and~\ref{app:details}).

Throughout these interactions, agents never see each other's reasoning or tool results; they share information only through messages they choose to send and read. A shared prompt describes the tools and the shared objective. It also gives general guidance on collaboration, suggesting useful kinds of messages while stating that no roles are assigned and that working alone is allowed (Appendix~\ref{app:prompts}).

\textbf{Formalization.} This setup can be modeled as a decentralized partially observable Markov decision process (Dec-POMDP) \citep{bernstein2013complexitydecentralizedcontrolmarkov}, the tuple $\langle \mathcal{I}, \mathcal{S}, \{\mathcal{A}_i\}, P, \{\Omega_i\}, O, R, H \rangle$. Here $\mathcal{I}=\{1,\dots,N\}$ is the set of agents. A state $s\in\mathcal{S}$ collects the agents' private task states, histories, message queues, and remaining budgets. The action space $\mathcal{A}_i$ of agent $i$ is generated text, including tool calls and messages. The transition function $P(s'\mid s,\mathbf{a})$ updates histories and budgets as text is generated, private task states when task tools execute, and message queues when agents send or read messages. The observation space $\Omega_i$ contains agent $i$'s private task input, tool responses, notifications, and the messages it reads, and the observation function $O(\mathbf{o}\mid s',\mathbf{a})$ determines what each agent receives after a joint action $\mathbf{a}$. The reward $R$ is the shared team return of Equation~\ref{eq:reward}, received by every agent, and the horizon $H$ is finite because each agent generates at most $L$ tokens. Agent $i$ acts on its own history $h_i$ of actions and observations, and the shared policy samples $a_i\sim\pi_\theta(\cdot\mid h_i)$.

\begin{figure}[t]
\centering
\sweepfigure{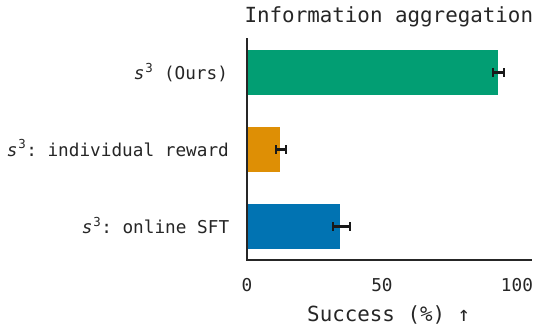}{\linewidth}
\caption{Bars show mean $\pm$ SE across six seeds on held-out problems at the training swarm size, $N=4$. Higher success is better. Ablations modify \method{}; solo policies train independently (Table~\ref{tab:conditions}).}
\label{fig:aggregation-controls}
\end{figure}

\subsection{Team reward}
\label{sec:reward}

Let $S\in\{0,1\}$ be binary team success, which is 1 if at least one agent reaches the correct final answer. The team reward is
\begin{equation}
R \;=\; S\left(1 - \lambda \min\!\left(1, \frac{\sum_{i=1}^{N} |\tau_i|}{N L}\right)\right),
\label{eq:reward}
\end{equation}
where $|\tau_i|$ is the number of tokens agent $i$ generated and $\lambda=0.25$. The cost is the fraction of the team's total allowance $NL$ that the team spent, so its effect is the same regardless of team size: a team of any size that spends half its allowance pays the same cost. We charge it only when the team succeeds. Charging failures would reward a failing team for quitting early. Among successful teams, it favors those that succeed with fewer tokens, for example by not sending redundant messages. Since $\lambda<1$, any success earns more than any failure.

The reward is the best outcome of the team, not the sum over its agents. A sum would make a larger team worth more simply because it is larger, and the loss below would weight teams by size. With the max, a team is rewarded only for what it achieves, so adding agents helps only if they contribute. An average over agents would instead reward each agent for being right itself, which encourages agents to solve the problem separately and then argue toward a consensus.

\subsection{Team-relative advantage}
\label{sec:advantage}

For each problem we sample $G=8$ independent team attempts, each with its own names, histories, and message board. With $R_g$ the reward of team $g$, every token generated by every member of team $g$ receives the advantage
\begin{equation}
A_g = \frac{R_g - \bar R}{\sigma_R + \epsilon},
\label{eq:advantage}
\end{equation}
where $\bar R$ and $\sigma_R$ are the mean and standard deviation of the $G$ team rewards. This is the group-relative advantage of GRPO \citep{shao2024deepseekmathpushinglimitsmathematical} with the team, not the single response, as the unit being compared. The comparison between independent teams on the same problem replaces per-agent credit assignment: behavior that separates successful teams from failed ones is reinforced, and behavior common to both cancels out.

We apply the loss only to tokens each agent generates. System prompts, tool results, notifications, and received messages provide context but are not training targets. Let $\mathcal T_g$ be the trained tokens of team $g$ and $\mathcal B$ the teams in a batch. For the decentralized exploration experiments we use IcePop \citep{lingteam2025everystep} with score centering to correct the gradient bias introduced by masking off-policy samples \citep{marek2026scorecenteringstabilizesoffpolicy}. We average over each team's generated tokens and then over teams, so teams with more agents or longer rollouts do not receive more weight. The policy-gradient term is
\begin{equation}
\mathcal L_{\rm PG}(\theta)
  = -\frac{1}{|\mathcal B|}\sum_{g\in\mathcal B}
     \frac{1}{|\mathcal T_g|}\sum_{t\in\mathcal T_g}m_t\,\rho_t\,A_g.
\label{eq:loss}
\end{equation}
Here $\rho_t=\pi_\theta(a_t\mid c_t)/\mu(a_t\mid c_t)$ compares the probability of token $a_t$ under the policy being trained with its probability under the policy $\mu$ that generated it. The mask $m_t=\mathbf 1[0.2\leq\rho_t\leq5.0]$ excludes tokens whose probabilities have changed too much, and is held fixed when differentiating.

Appendix~\ref{app:extended-methods} describes the training configuration, baselines, evaluation protocol, and score-centering correction in detail.

\begin{figure*}[t]
\centering
\sweepfigure{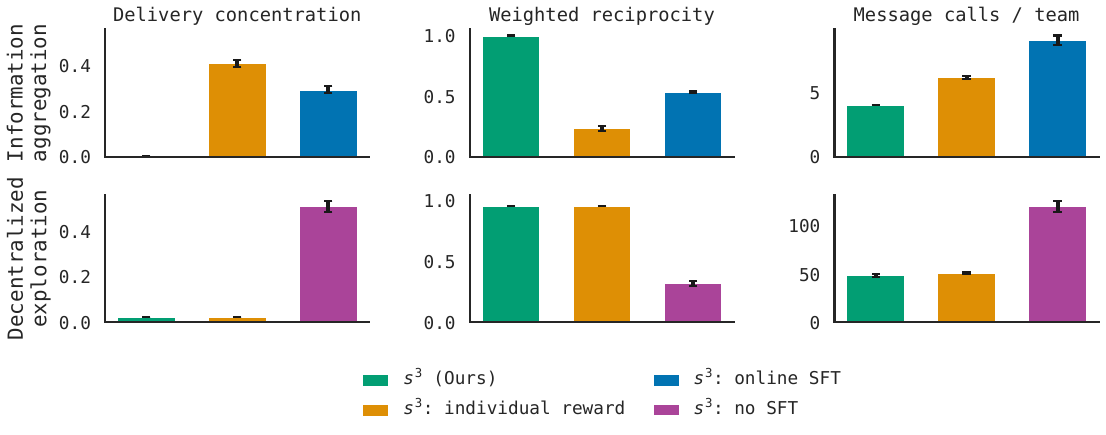}{0.80\linewidth}
\caption{Communication structure in final-policy evaluation episodes at $N=4$: information aggregation above, decentralized exploration below. Only conditions with messaging enabled are shown; learner names follow Table~\ref{tab:conditions}. The first two columns measure whether a few agents dominate sending and how much communication flows in both directions; the third counts message calls per team. Broadcasts count once as calls and once per recipient as deliveries. Bars show mean $\pm$ SE across six seeds. The main text reports the message-structure and room-overlap comparisons; Appendix Table~\ref{tab:key-tests} summarizes selected tests.}
\label{fig:graphs}
\end{figure*}

\subsection{Training and evaluation}
Information aggregation uses Qwen3-0.6B; decentralized exploration uses Qwen3-1.7B. Swarms train with four agents for 200 RL updates. Exploration learners start from 50 updates of supervised fine-tuning (SFT), except in the ablation without SFT. Solo learners receive the same number of agent trajectories per update and the same per-agent token allowance. We evaluate on 32 held-out problems with six seeds and one attempt per problem per seed, using populations of $N\in\{1,2,4,6,8,12,16\}$ without retraining. Token efficiency is solve rate divided by the mean fraction of the population's token allowance used, including failed attempts. Appendix~\ref{app:extended-methods} gives the full protocol.

\section{Tasks and why coordination is hard to scale}
\label{sec:tasks}

\subsection{Coordination vs. parallel exploration}
\label{sec:scaling}

Communication can help a swarm combine information that its members hold privately or coordinate the search for information they have yet to discover \citep{foerster2016learningcommunicatedeepmultiagent,sukhbaatar2016learningmultiagentcommunicationbackpropagation}. These settings place different demands on a policy as the number of agents grows. We train with four agents and evaluate with up to sixteen to test both.

\textbf{Information aggregation.} When each agent holds information that the others need, solving the task requires coordination: at least one agent must collect and correctly combine the distributed information. Adding independent agents cannot substitute for this exchange. Suppose each agent holds one piece of information and each of the $N-1$ transfers needed by a recipient succeeds independently with probability $q<1$. A transfer includes sending, reading, and correctly recording the information. That recipient has all the information with probability $q^{N-1}$, and a union bound over possible recipients gives
\begin{equation}
P_{\mathrm{coord}}(N) \le \min\!\left(1, N q^{N-1}\right),
\label{eq:coord}
\end{equation}
which eventually decays exponentially in $N$ at fixed $q$. Sustaining success as swarms grow therefore requires increasingly reliable information exchange. The problem also becomes more complex as the swarm grows: an agent's inbox can be in $2^{N-1}$ states depending on who has reported, 8 with four agents and 32,768 with sixteen. Agents must learn a rule such as ``wait for $N-1$ distinct senders'' rather than a behavior for each state. If every agent broadcasts once, the swarm exchanges $N(N-1)$ message deliveries in total. Each agent receives $N-1$ messages, which take up context space and leave less room for its own reasoning and task history (Appendix~\ref{app:coordination}).

\textbf{Decentralized exploration.} Here, any agent can find a solution independently, but communication can help agents divide the search and use one another's discoveries. If each independent attempt succeeds with probability $p$, best-of-$N$ sampling succeeds with probability
\begin{equation}
P_{\mathrm{search}}(N) = 1-(1-p)^N,
\label{eq:search}
\end{equation}
which rises with $N$. Unlike information aggregation, adding agents does not require an independent search policy to manage more incoming information. Communication must improve on this growing baseline by reducing repeated exploration or directing agents toward useful discoveries. Generalizing to larger swarms thus tests whether agents can maintain reliable information exchange in aggregation and coordinate exploration well enough to improve on independent search.

\subsection{Task implementations}

\textbf{Information aggregation.} Each of the $N$ agents receives one private integer from 1 to 99, written in English (e.g. ``forty-two''). The swarm must return all $N$ numbers in ascending order, keeping equal numbers from different agents as separate entries. It succeeds if any agent submits the full, correct list. Ordering the numbers is simple once they are available; the challenge is to share one's own number, read the others, track who owns which value so repeated messages do not create duplicates, and submit only after hearing from everyone. The shared prompt explains these requirements and gives example message and read calls (Appendix~\ref{app:prompts}). Agents must learn to carry out these instructions reliably as messages arrive in an unpredictable order.

\begin{figure}[t]
\centering
\sweepfigure{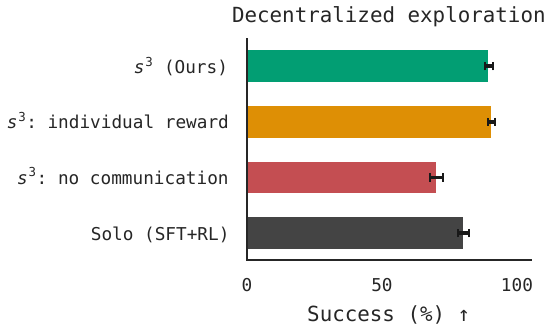}{\linewidth}
\caption{Bars show mean $\pm$ SE across six seeds on held-out problems at the training swarm size, $N=4$. Higher success is better. Ablations modify \method{}; solo policies train independently (Table~\ref{tab:conditions}).}
\label{fig:exploration-controls}
\end{figure}

\begin{figure*}[t]
\centering
\sweepfigure{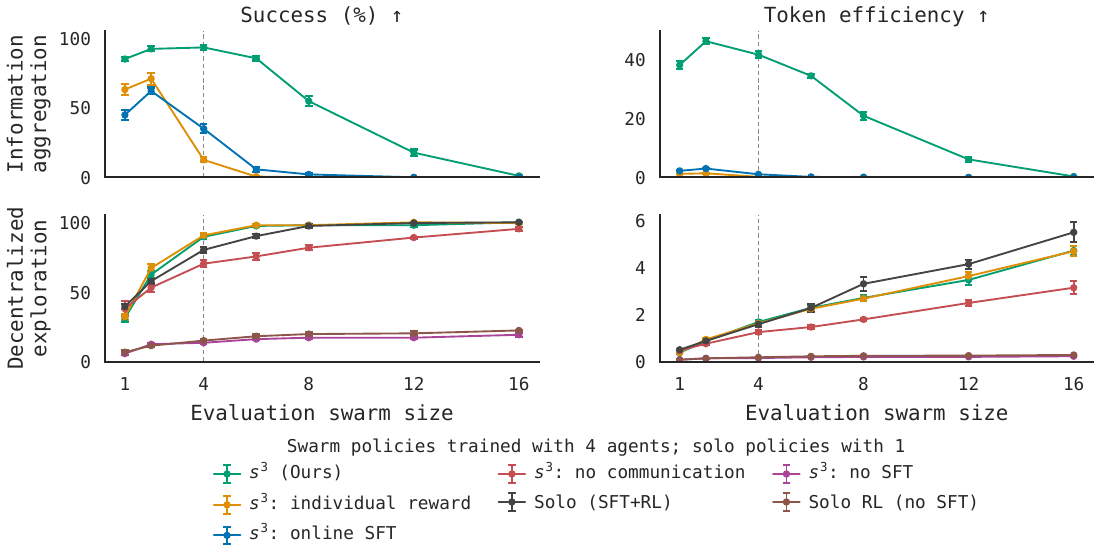}{0.8\linewidth}
\caption{Held-out evaluation across swarm sizes: information aggregation above, decentralized exploration below; success on the left and token efficiency on the right. Higher is better. Swarms train with four agents (dashed line); solo learners train with one. Other population sizes test generalization without retraining. Token efficiency is solve rate divided by the mean fraction of the population token allowance used, failures included. Points and error bars show mean $\pm$ SE across six seeds. Names follow Table~\ref{tab:conditions}.}
\label{fig:scaling-comparison}
\label{fig:aggregation-scaling}
\label{fig:exploration-scaling}
\end{figure*}

\textbf{Decentralized exploration.} Each agent explores an unknown tree of rooms. Each junction has $b=3$ doors leading deeper into the tree, with corridors requiring ten moves between successive junctions or terminal rooms, down to depth $d$. One of the $b^d$ deepest terminal rooms is the exit. Passages can be traversed in either direction. Room names are random, so they carry no information. Each room lists its doors in an order fixed at generation and shared by all agents, so agents with similar search habits retrace the same paths. Every agent starts at the same room in its own copy of the tree and uses \texttt{move(room)} and \texttt{look()} to observe its current room, its doors, and whether it is a dead end. Moving never moves another agent. The first agent to reach the exit solves the task for the swarm and ends the episode. At depth three, the tree has 391 rooms and 27 terminal rooms; a full depth-first search can take up to 780 moves, usually beyond what a reasoning model can afford within its 8,192-token allowance. Agents can reduce repeated work by claiming branches or reporting exhausted ones, but must learn how to use these messages to coordinate their search.

\section{Results}
\label{sec:results}

We first examine whether swarms learn to coordinate in each task, then test generalization to larger numbers of agents and the contribution of each training component. Results use held-out problems and six evaluation seeds. Statistical comparisons use paired, two-sided $t$ tests with Bonferroni correction for multiple comparisons; we report paired effect sizes as $d_z$. Appendix~\ref{app:extended-methods} and Appendix Table~\ref{tab:key-tests} give the test definitions and detailed comparisons.

\subsection{RQ1: Can swarms learn to aggregate private information?}

Figure~\ref{fig:aggregation-controls} shows that \method{} learns reliable information aggregation, substantially outperforming both controls. It reaches 93\% success, compared with 35\% for \method{}: online SFT on successful swarms and 13\% for \method{}: individual reward (both corrected $p<0.001$; $d_z=5.72$ and $11.15$, respectively). In the individual-reward variant, agents still communicate within a swarm, but each is rewarded only when its own submitted list is correct; helping another agent answer earns no reward. Online SFT instead imitates the actions of successful swarms. Both controls receive the same collection budget as \method{}, so additional interaction or successful examples alone do not explain its advantage.

\method{} agents also send fewer messages than the controls. In Figure~\ref{fig:graphs} (top), each \method{} agent broadcasts its number about once, and all agents send about equally. The controls send more messages, but a few agents send most of them while the rest send little. Section~\ref{sec:analysis} traces how this behavior develops during training.

\subsection{RQ2: Does learned coordination improve decentralized exploration?}

Figure~\ref{fig:exploration-controls} shows that communicating swarms also benefit when each agent could solve the task alone. Both swarm and solo policies begin from demonstrations of valid exploration, with messages included only in the swarm demonstrations. Communication already raises success before RL, and RL then improves both the communicating swarm's success and its token efficiency (Appendix Figure~\ref{fig:training-exploration-sft}; Appendix Table~\ref{tab:initial-final}).

The ablations show that communication contributes to this gain. Removing messages while retaining the shared reward lowers four-agent success by 19 percentage points (corrected $p<0.01$, $d_z=2.21$). Communicating agents also explore fewer of the same rooms: their overlap is nearly half that of the communication-disabled control (Appendix Table~\ref{tab:key-tests}). They use discoveries to divide the search, rather than simply adding more independent attempts. At four agents, \method{} also has higher success than solo-trained best-of-four, which is significant before Bonferroni correction but not after (Appendix Table~\ref{tab:key-tests}).

\subsection{RQ3: Do these gains extend to larger numbers of agents?}

Figure~\ref{fig:scaling-comparison} evaluates the same policies with more agents than the four used in training. In information aggregation, \method{} retains useful success after the controls largely fail: at eight agents it succeeds more than twenty times as often as online SFT. Its advantage over both controls persists through twelve agents (all corrected $p<0.01$, $d_z\geq3.24$; Appendix Table~\ref{tab:key-tests}). All methods struggle at sixteen. The learned exchange therefore transfers beyond the training size, but does not eliminate the growing difficulty of collecting and tracking everyone's information.

In decentralized exploration, independent best-of-$N$ already improves as agents are added. Communicating swarms reach high success sooner: four agents nearly match six solo-trained agents, and six nearly match eight. At six agents, \method{} exceeds best-of-six by about seven percentage points (corrected $p<0.05$, $d_z=1.71$). This gain comes without a statistically detectable decrease in token efficiency after correction. The same is true of the efficiency comparisons at the larger evaluated swarm sizes (Appendix Table~\ref{tab:paired-maze}; Appendix Figure~\ref{fig:paired-gains}). At the largest sizes, independent search also approaches perfect success, so communication's main benefit is reaching that performance with fewer agents.

\begin{figure*}[t]
\centering
\includegraphics[width=\textwidth]{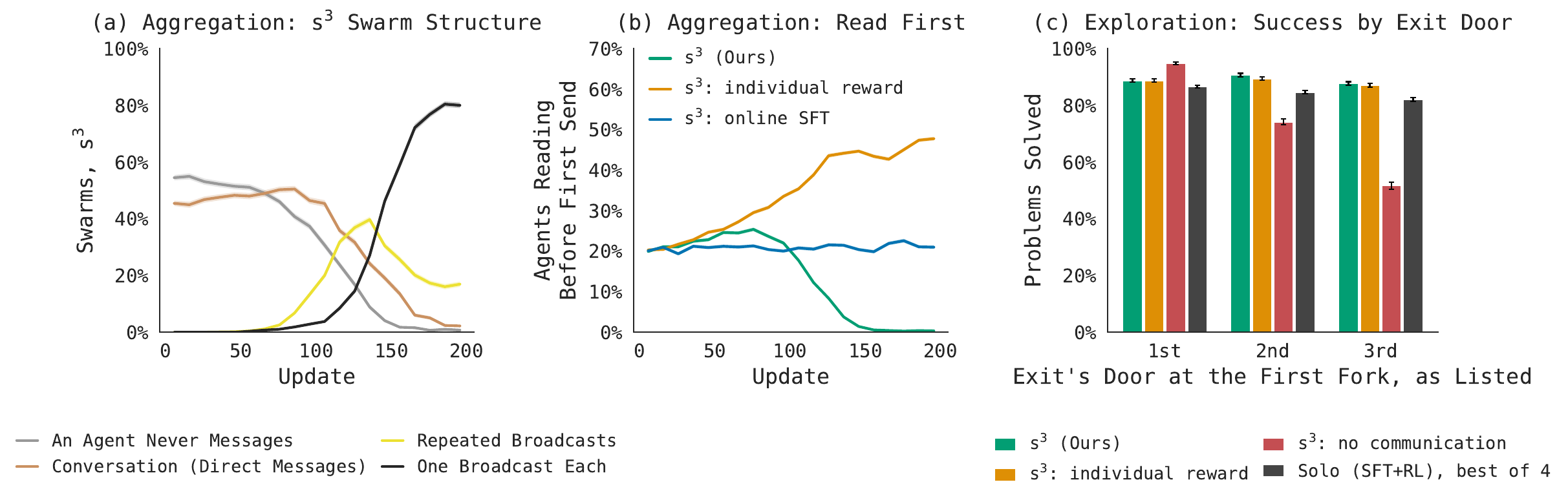}
\caption{How coordination changes during training. (a) Information aggregation: swarms progress from incomplete information sharing, through repeated broadcasts, toward one broadcast per agent. Categories distinguish incomplete participation (at least one agent never sends) from swarms where everyone sends: with direct messages, with broadcasts only but some repeated, or with one broadcast per agent. (b) Agents that read incoming messages before sharing their own information. (c) Decentralized exploration: success according to which branch at the first junction contains the exit. Communication-disabled agents favor the first-listed branch; communicating swarms succeed more consistently across exit locations. Bands and bars show mean $\pm$ SE across groups of attempts on the same problem in ten-update bins (a, b) or across search trees (c).}
\label{fig:analysis}
\end{figure*}

\subsection{RQ4: Which ingredients of \method{} drive its gains?}

We first test whether decentralized exploration needs the demonstrations that both swarm and solo policies start from. Without SFT, both swarm and solo learners remain much less successful, and the four-agent swarm has no significant advantage over solo-trained best-of-four (uncorrected $p>0.05$, $d_z=-0.48$). Starting from the base model, RL raises swarm success only modestly. Demonstrations instead start both swarm and solo policies at much higher success, and RL then keeps improving the swarm (Appendix Figures~\ref{fig:exploration-initial-final}, \ref{fig:training-exploration-sft}, and~\ref{fig:training-exploration-no-sft}; Appendix Table~\ref{tab:initial-final}).

One explanation for these gains is that agents can communicate, not that they are trained to help each other. To test this, we train a control in which agents still work as a swarm and can message each other, but each agent is rewarded only for solving the task itself: for reaching the exit in decentralized exploration, or for submitting a correct list in information aggregation. In decentralized exploration, this control performs about as well as shared-reward training, with no significant success difference (uncorrected $p>0.05$, $d_z=-0.28$), and it also outperforms solo-trained best-of-four (corrected $p<0.01$, $d_z=2.21$). This is expected: any agent can reach the exit alone, so reports from teammates help each agent earn its own reward. In information aggregation, however, an agent gains nothing by sending its own number, since that only helps its teammates answer. Individual reward fails there (Figure~\ref{fig:aggregation-controls}), and agents learn to stay silent (Section~\ref{sec:analysis}).

Together, the ablations show why the shared objective is useful across tasks. Individual rewards support effective exploration when an agent can earn its own reward by finding the exit, but perform poorly when success depends on teammates sharing private information. Shared rewards support both forms of coordination, so swarms can adapt even when we do not know the structure of a problem in advance.

\section{How organization emerges in decentralized swarms}
\label{sec:analysis}

In this section, we analyze the swarms' training traces to see how agents learn to share information and divide their search over the course of RL. Appendix~\ref{app:analysis} describes the data, sampling, and statistical analysis in more detail.

\textbf{Information aggregation swarms learn to share once and share early.} We first group each information aggregation swarm by how its agents message one another and follow these groups over training (Figure~\ref{fig:analysis}a). Early on, about half of swarms include an agent that never sends its number, and most of the rest exchange direct messages. Partway through training, swarms switch to broadcasts, at first repeating them, and by the end about 80\% settle on one broadcast per agent. These learners start from the base model without demonstrations, so this exchange emerges from RL; the prompt only shows an example broadcast. Agents also learn to share before they read: the share of \method{} agents that check their inbox before sending their own number falls to nearly zero (Figure~\ref{fig:analysis}b).

Early in training, while few swarms use one broadcast per agent, those that do succeed about 30 percentage points more often than swarms that converse through direct messages on the same problem with the same policy. The shared reward therefore favors this exchange as training continues (Appendix Figure~\ref{fig:app-assoc}).

\textbf{Under individual rewards, agents stop sharing.} We also follow the individual-reward swarms from RQ4, where each agent is rewarded only for its own answer. Here agents move the other way: more and more of them read their teammates' numbers before sending their own, and many eventually never send at all (Figure~\ref{fig:analysis}b). This is a social dilemma, much like a public goods game \citep{Olson1974-wt, leibo2017multiagentreinforcementlearningsequential}: every agent would have a better chance of answering correctly if all of them shared, but an agent's own number only helps its teammates, so each one is better off waiting for the others to share first. When every agent waits, no one shares. A shared reward removes the dilemma, because an agent gains whenever any teammate answers correctly. Decentralized exploration has no such conflict, which is why individual-reward swarms still do well there.

\textbf{Communicating explorers split up the search.} In decentralized exploration, every agent sees a room's doors listed in the same order, so agents with similar habits tend to take the same first door. If they all start down the same branch, they find the exit quickly only when it lies down that branch. To test this, we sort the held-out search trees by whether the exit lies behind the first, second, or third door listed at the first junction, and compare success across these three groups (Figure~\ref{fig:analysis}c). Swarms that cannot communicate succeed less often when the exit is behind the second door and much less often when it is behind the third. \method{} swarms succeed at about the same rate wherever the exit is, because agents claim different branches and report the ones they have finished. These findings suggest that \method{} trains swarms to use messages to divide the search effectively, rather than to communicate arbitrarily (Appendix Figures~\ref{fig:app-maze-problems} and~\ref{fig:app-maze-comm}).

\section{Discussion and next steps}
\label{sec:discussion}

These experiments show that language-model swarms can learn effective communication and coordination through cooperative MARL. The asynchronous infrastructure lets agents exchange discoveries while their partners are still working, and the shared reward trains those exchanges alongside task actions. In information aggregation, swarms learn an efficient exchange that continues to work with more agents than were used in training. Communicating swarms improve decentralized exploration at the same population size, and smaller swarms nearly match larger solo-trained best-of-$N$ populations.

Our experiments are limited to two tasks, small models, and one training run per condition, so additional runs may change some of our findings. The analysis of the swarms' behavior over the course of training points to a next step: agents sometimes submit a wrong answer despite having all the information, repeat a teammate's search, or trust an unverified report. Training agents to track what their partners know and whether their reports are justified could make larger swarms more reliable without more communication, and broader tasks and larger models will test how far these findings extend.

The main result is that effective collaboration can be learned in the same process as task solving. With a shared goal and infrastructure that allows agents to act asynchronously, useful ways of exchanging information and dividing work develop during training. Their transfer to larger populations shows how training a small swarm can support scaling at inference.

\FloatBarrier
\bibliographystyle{sn-nature}
\bibliography{references,workshop}

\begin{thebibliography}{10}
\expandafter\ifx\csname url\endcsname\relax
  \def\url#1{\burl{#1}}\fi
\expandafter\ifx\csname urlprefix\endcsname\relax\def\urlprefix{URL }\fi
\providecommand{\bibinfo}[2]{#2}
\providecommand{\eprint}[2][]{\url{#2}}
\providecommand{\doi}[1]{\url{https://doi.org/#1}}
\bibcommenthead

\bibitem[{Humphrey(1988)}]{humphrey1976social}
\bibinfo{author}{Humphrey, N.~K.}
\newblock \bibinfo{title}{ in \textit{The social function of intellect}} \emph{\bibinfo{booktitle}{Machiavellian Intelligence: Social Expertise and the Evolution of Intellect in Monkeys, Apes, and Humans}} (eds  \bibinfo{editor}{Byrne, R.~W.} \& \bibinfo{editor}{Whiten, A.}) \bibinfo{pages}{13--26} (\bibinfo{publisher}{Clarendon Press/Oxford University Press}, \bibinfo{address}{New York, NY, US}, \bibinfo{year}{1988}).

\bibitem[{Dunbar(2009)}]{dunbar1998social}
\bibinfo{author}{Dunbar, R. I.~M.}
\newblock \bibinfo{title}{The social brain hypothesis and its implications for social evolution}.
\newblock \emph{\bibinfo{journal}{Ann. Hum. Biol.}} \textbf{\bibinfo{volume}{36}}, \bibinfo{pages}{562--572} (\bibinfo{year}{2009}).

\bibitem[{Henrich(2016)}]{henrich2016secret}
\bibinfo{author}{Henrich, J.}
\newblock \emph{\bibinfo{title}{The secret of our success}}  (\bibinfo{publisher}{Princeton University Press}, \bibinfo{address}{Princeton, NJ}, \bibinfo{year}{2016}).

\bibitem[{OpenAI(2026)}]{openaiNavierStokesMillennium}
\bibinfo{author}{OpenAI}.
\newblock \bibinfo{title}{{O}n the {N}avier–{S}tokes {M}illennium {P}rize {P}roblem}.
\newblock \bibinfo{howpublished}{\url{https://openai.com/index/navier-stokes-solution/}} (\bibinfo{year}{2026}).
\newblock \bibinfo{note}{[Accessed 18-09-2026]}.

\bibitem[{Team et~al.(2026)}]{kimiteam2026kimik25visualagentic}
\bibinfo{author}{Team, K.} \emph{et~al.}
\newblock \bibinfo{title}{Kimi k2.5: Visual agentic intelligence} (\bibinfo{year}{2026}).
\newblock \urlprefix\url{https://arxiv.org/abs/2602.02276}.
\newblock \bibinfo{eprint}{{\href{https://arxiv.org/abs/2602.02276}{{arXiv:2602.02276}}}}.

\bibitem[{Park et~al.(2026)}]{park2026scalingdiscoverytesttimecommunication}
\bibinfo{author}{Park, J.}, \bibinfo{author}{Kontonis, V.}, \bibinfo{author}{Garg, S.}, \bibinfo{author}{Krishnamurthy, A.} \& \bibinfo{author}{Papailiopoulos, D.}
\newblock \bibinfo{title}{Scaling discovery through test-time communication} (\bibinfo{year}{2026}).
\newblock \urlprefix\url{https://arxiv.org/abs/2609.21032}.
\newblock \bibinfo{eprint}{{\href{https://arxiv.org/abs/2609.21032}{{arXiv:2609.21032}}}}.

\bibitem[{Wu et~al.(2023)}]{wu2023autogenenablingnextgenllm}
\bibinfo{author}{Wu, Q.} \emph{et~al.}
\newblock \bibinfo{title}{Autogen: Enabling next-gen llm applications via multi-agent conversation} (\bibinfo{year}{2023}).
\newblock \urlprefix\url{https://arxiv.org/abs/2308.08155}.
\newblock \bibinfo{eprint}{{\href{https://arxiv.org/abs/2308.08155}{{arXiv:2308.08155}}}}.

\bibitem[{Fourney et~al.(2024)}]{fourney2024magenticonegeneralistmultiagentsolving}
\bibinfo{author}{Fourney, A.} \emph{et~al.}
\newblock \bibinfo{title}{Magentic-one: A generalist multi-agent system for solving complex tasks} (\bibinfo{year}{2024}).
\newblock \urlprefix\url{https://arxiv.org/abs/2411.04468}.
\newblock \bibinfo{eprint}{{\href{https://arxiv.org/abs/2411.04468}{{arXiv:2411.04468}}}}.

\bibitem[{Hadfield et~al.(2025)}]{anthropicBuiltMultiagent}
\bibinfo{author}{Hadfield, J.} \emph{et~al.}
\newblock \bibinfo{title}{{H}ow we built our multi-agent research system}.
\newblock \bibinfo{howpublished}{\url{https://www.anthropic.com/research/multi-agent-research-system}} (\bibinfo{year}{2025}).
\newblock \bibinfo{note}{[Accessed 03-11-2026]}.

\bibitem[{Du et~al.(2023)}]{du2023improvingfactualityreasoninglanguage}
\bibinfo{author}{Du, Y.}, \bibinfo{author}{Li, S.}, \bibinfo{author}{Torralba, A.}, \bibinfo{author}{Tenenbaum, J.~B.} \& \bibinfo{author}{Mordatch, I.}
\newblock \bibinfo{title}{Improving factuality and reasoning in language models through multiagent debate} (\bibinfo{year}{2023}).
\newblock \urlprefix\url{https://arxiv.org/abs/2305.14325}.
\newblock \bibinfo{eprint}{{\href{https://arxiv.org/abs/2305.14325}{{arXiv:2305.14325}}}}.

\bibitem[{Hong et~al.(2024)}]{hong2024metagptmetaprogrammingmultiagent}
\bibinfo{author}{Hong, S.} \emph{et~al.}
\newblock \bibinfo{title}{Metagpt: Meta programming for a multi-agent collaborative framework} (\bibinfo{year}{2024}).
\newblock \urlprefix\url{https://arxiv.org/abs/2308.00352}.
\newblock \bibinfo{eprint}{{\href{https://arxiv.org/abs/2308.00352}{{arXiv:2308.00352}}}}.

\bibitem[{Cemri et~al.(2025)}]{cemri2025multiagentllmsystemsfail}
\bibinfo{author}{Cemri, M.} \emph{et~al.}
\newblock \bibinfo{title}{Why do multi-agent llm systems fail?} (\bibinfo{year}{2025}).
\newblock \urlprefix\url{https://arxiv.org/abs/2503.13657}.
\newblock \bibinfo{eprint}{{\href{https://arxiv.org/abs/2503.13657}{{arXiv:2503.13657}}}}.

\bibitem[{Boyd and Richerson(1985)}]{Boyd1985-gr}
\bibinfo{author}{Boyd, R.} \& \bibinfo{author}{Richerson, P.~J.}
\newblock \emph{\bibinfo{title}{Culture and the evolutionary process}}  (\bibinfo{publisher}{University of Chicago Press}, \bibinfo{address}{Chicago, IL}, \bibinfo{year}{1985}).

\bibitem[{Hayek(1945)}]{hayek}
\bibinfo{author}{Hayek, F.~A.}
\newblock \bibinfo{title}{The use of knowledge in society}.
\newblock \emph{\bibinfo{journal}{The American Economic Review}} \textbf{\bibinfo{volume}{35}}, \bibinfo{pages}{519--530} (\bibinfo{year}{1945}).
\newblock \urlprefix\url{http://www.jstor.org/stable/1809376}.

\bibitem[{Sutton(2019)}]{incompleteideasBitterLesson}
\bibinfo{author}{Sutton, R.}
\newblock \bibinfo{title}{The bitter lesson}.
\newblock \bibinfo{howpublished}{\url{http://www.incompleteideas.net/{I}nc{I}deas/{B}itter{L}esson.html}} (\bibinfo{year}{2019}).
\newblock \bibinfo{note}{[Accessed 03-11-2026]}.

\bibitem[{Silver et~al.(2021)}]{Silver2021-af}
\bibinfo{author}{Silver, D.}, \bibinfo{author}{Singh, S.}, \bibinfo{author}{Precup, D.} \& \bibinfo{author}{Sutton, R.~S.}
\newblock \bibinfo{title}{Reward is enough}.
\newblock \emph{\bibinfo{journal}{Artif. Intell.}} \textbf{\bibinfo{volume}{299}}, \bibinfo{pages}{103535} (\bibinfo{year}{2021}).

\bibitem[{OpenAI et~al.(2024)}]{openai2026openaio1card}
\bibinfo{author}{OpenAI} \emph{et~al.}
\newblock \bibinfo{title}{Openai o1 system card} (\bibinfo{year}{2024}).
\newblock \urlprefix\url{https://arxiv.org/abs/2412.16720}.
\newblock \bibinfo{eprint}{{\href{https://arxiv.org/abs/2412.16720}{{arXiv:2412.16720}}}}.

\bibitem[{Guo et~al.(2025)}]{Guo_2025}
\bibinfo{author}{Guo, D.} \emph{et~al.}
\newblock \bibinfo{title}{Deepseek-r1 incentivizes reasoning in llms through reinforcement learning}.
\newblock \emph{\bibinfo{journal}{Nature}} \textbf{\bibinfo{volume}{645}}, \bibinfo{pages}{633–638} (\bibinfo{year}{2025}).
\newblock \urlprefix\url{http://dx.doi.org/10.1038/s41586-025-09422-z}.

\bibitem[{Brown et~al.(2024)}]{brown2024largelanguagemonkeysscaling}
\bibinfo{author}{Brown, B.} \emph{et~al.}
\newblock \bibinfo{title}{Large language monkeys: Scaling inference compute with repeated sampling} (\bibinfo{year}{2024}).
\newblock \urlprefix\url{https://arxiv.org/abs/2407.21787}.
\newblock \bibinfo{eprint}{{\href{https://arxiv.org/abs/2407.21787}{{arXiv:2407.21787}}}}.

\bibitem[{Wang et~al.(2023)}]{wang2023selfconsistencyimproveschainthought}
\bibinfo{author}{Wang, X.} \emph{et~al.}
\newblock \bibinfo{title}{Self-consistency improves chain of thought reasoning in language models} (\bibinfo{year}{2023}).
\newblock \urlprefix\url{https://arxiv.org/abs/2203.11171}.
\newblock \bibinfo{eprint}{{\href{https://arxiv.org/abs/2203.11171}{{arXiv:2203.11171}}}}.

\bibitem[{Snell et~al.(2024)}]{snell2024scalingllmtesttimecompute}
\bibinfo{author}{Snell, C.}, \bibinfo{author}{Lee, J.}, \bibinfo{author}{Xu, K.} \& \bibinfo{author}{Kumar, A.}
\newblock \bibinfo{title}{Scaling llm test-time compute optimally can be more effective than scaling model parameters} (\bibinfo{year}{2024}).
\newblock \urlprefix\url{https://arxiv.org/abs/2408.03314}.
\newblock \bibinfo{eprint}{{\href{https://arxiv.org/abs/2408.03314}{{arXiv:2408.03314}}}}.

\bibitem[{Zarov et~al.(2026)}]{generalreasoning2026avalonm1}
\bibinfo{author}{Zarov, I.} \emph{et~al.}
\newblock \bibinfo{title}{Avalon-m1: Reinforcement learning from collective rewards} (\bibinfo{year}{2026}).
\newblock \urlprefix\url{https://gr.inc/avalonm1.pdf}.

\bibitem[{Jha et~al.(2026)}]{jha2026solosociallearningcharacterizing}
\bibinfo{author}{Jha, K.}, \bibinfo{author}{Kleiman-Weiner, M.} \& \bibinfo{author}{Jaques, N.}
\newblock \bibinfo{title}{From solo to social learning: Characterizing recursive social improvement in llms} (\bibinfo{year}{2026}).
\newblock \urlprefix\url{https://arxiv.org/abs/2609.38516}.
\newblock \bibinfo{eprint}{{\href{https://arxiv.org/abs/2609.38516}{{arXiv:2609.38516}}}}.

\bibitem[{Stanley and Lehman(2015)}]{Stanley2015-nl}
\bibinfo{author}{Stanley, K.~O.} \& \bibinfo{author}{Lehman, J.}
\newblock \emph{\bibinfo{title}{Why greatness cannot be planned: The myth of the objective}}  (\bibinfo{publisher}{Springer International Publishing}, \bibinfo{year}{2015}).

\bibitem[{Li et~al.(2023)}]{li2023camelcommunicativeagentsmind}
\bibinfo{author}{Li, G.}, \bibinfo{author}{Hammoud, H. A. A.~K.}, \bibinfo{author}{Itani, H.}, \bibinfo{author}{Khizbullin, D.} \& \bibinfo{author}{Ghanem, B.}
\newblock \bibinfo{title}{Camel: Communicative agents for "mind" exploration of large language model society} (\bibinfo{year}{2023}).
\newblock \urlprefix\url{https://arxiv.org/abs/2303.17760}.
\newblock \bibinfo{eprint}{{\href{https://arxiv.org/abs/2303.17760}{{arXiv:2303.17760}}}}.

\bibitem[{Qian et~al.(2024)}]{qian2024chatdevcommunicativeagentssoftware}
\bibinfo{author}{Qian, C.} \emph{et~al.}
\newblock \bibinfo{title}{Chatdev: Communicative agents for software development} (\bibinfo{year}{2024}).
\newblock \urlprefix\url{https://arxiv.org/abs/2307.07924}.
\newblock \bibinfo{eprint}{{\href{https://arxiv.org/abs/2307.07924}{{arXiv:2307.07924}}}}.

\bibitem[{Chen et~al.(2023)}]{chen2023agentversefacilitatingmultiagentcollaboration}
\bibinfo{author}{Chen, W.} \emph{et~al.}
\newblock \bibinfo{title}{Agentverse: Facilitating multi-agent collaboration and exploring emergent behaviors} (\bibinfo{year}{2023}).
\newblock \urlprefix\url{https://arxiv.org/abs/2308.10848}.
\newblock \bibinfo{eprint}{{\href{https://arxiv.org/abs/2308.10848}{{arXiv:2308.10848}}}}.

\bibitem[{Wang et~al.(2024)}]{wang2024mixtureofagentsenhanceslargelanguage}
\bibinfo{author}{Wang, J.}, \bibinfo{author}{Wang, J.}, \bibinfo{author}{Athiwaratkun, B.}, \bibinfo{author}{Zhang, C.} \& \bibinfo{author}{Zou, J.}
\newblock \bibinfo{title}{Mixture-of-agents enhances large language model capabilities} (\bibinfo{year}{2024}).
\newblock \urlprefix\url{https://arxiv.org/abs/2406.04692}.
\newblock \bibinfo{eprint}{{\href{https://arxiv.org/abs/2406.04692}{{arXiv:2406.04692}}}}.

\bibitem[{Zhang et~al.(2025)}]{zhang2025aflowautomatingagenticworkflow}
\bibinfo{author}{Zhang, J.} \emph{et~al.}
\newblock \bibinfo{title}{Aflow: Automating agentic workflow generation} (\bibinfo{year}{2025}).
\newblock \urlprefix\url{https://arxiv.org/abs/2410.10762}.
\newblock \bibinfo{eprint}{{\href{https://arxiv.org/abs/2410.10762}{{arXiv:2410.10762}}}}.

\bibitem[{Hu et~al.(2025)}]{hu2025automateddesignagenticsystems}
\bibinfo{author}{Hu, S.}, \bibinfo{author}{Lu, C.} \& \bibinfo{author}{Clune, J.}
\newblock \bibinfo{title}{Automated design of agentic systems} (\bibinfo{year}{2025}).
\newblock \urlprefix\url{https://arxiv.org/abs/2408.08435}.
\newblock \bibinfo{eprint}{{\href{https://arxiv.org/abs/2408.08435}{{arXiv:2408.08435}}}}.

\bibitem[{Zhang et~al.(2025)}]{zhang2025multiagentarchitecturesearchagentic}
\bibinfo{author}{Zhang, G.} \emph{et~al.}
\newblock \bibinfo{title}{Multi-agent architecture search via agentic supernet} (\bibinfo{year}{2025}).
\newblock \urlprefix\url{https://arxiv.org/abs/2502.04180}.
\newblock \bibinfo{eprint}{{\href{https://arxiv.org/abs/2502.04180}{{arXiv:2502.04180}}}}.

\bibitem[{Zhang et~al.(2025)}]{zhang2025gdesignerarchitectingmultiagentcommunication}
\bibinfo{author}{Zhang, G.} \emph{et~al.}
\newblock \bibinfo{title}{G-designer: Architecting multi-agent communication topologies via graph neural networks} (\bibinfo{year}{2025}).
\newblock \urlprefix\url{https://arxiv.org/abs/2410.11782}.
\newblock \bibinfo{eprint}{{\href{https://arxiv.org/abs/2410.11782}{{arXiv:2410.11782}}}}.

\bibitem[{Zhuge et~al.(2024)}]{zhuge2024languageagentsoptimizablegraphs}
\bibinfo{author}{Zhuge, M.} \emph{et~al.}
\newblock \bibinfo{title}{Language agents as optimizable graphs} (\bibinfo{year}{2024}).
\newblock \urlprefix\url{https://arxiv.org/abs/2402.16823}.
\newblock \bibinfo{eprint}{{\href{https://arxiv.org/abs/2402.16823}{{arXiv:2402.16823}}}}.

\bibitem[{Zhou et~al.(2025)}]{zhou2025resorewarddrivenselforganizingllmbased}
\bibinfo{author}{Zhou, H.} \emph{et~al.}
\newblock \bibinfo{title}{Reso: A reward-driven self-organizing llm-based multi-agent system for reasoning tasks} (\bibinfo{year}{2025}).
\newblock \urlprefix\url{https://arxiv.org/abs/2503.02390}.
\newblock \bibinfo{eprint}{{\href{https://arxiv.org/abs/2503.02390}{{arXiv:2503.02390}}}}.

\bibitem[{Cao et~al.(2026)}]{cao2026evolvingagentteams}
\bibinfo{author}{Cao, S.}, \bibinfo{author}{Mao, Z.}, \bibinfo{author}{Li, D.}, \bibinfo{author}{Gonzalez, J.~E.} \& \bibinfo{author}{Stoica, I.}
\newblock \bibinfo{title}{Evolving agent teams} (\bibinfo{year}{2026}).
\newblock \urlprefix\url{https://arxiv.org/abs/2610.36848}.
\newblock \bibinfo{eprint}{{\href{https://arxiv.org/abs/2610.36848}{{arXiv:2610.36848}}}}.

\bibitem[{Pappu et~al.(2026)}]{pappu2026selforganizingagentteamslearn}
\bibinfo{author}{Pappu, A.} \emph{et~al.}
\newblock \bibinfo{title}{Self-organizing agent teams learn to reason together} (\bibinfo{year}{2026}).
\newblock \urlprefix\url{https://arxiv.org/abs/2609.22682}.
\newblock \bibinfo{eprint}{{\href{https://arxiv.org/abs/2609.22682}{{arXiv:2609.22682}}}}.

\bibitem[{Park et~al.(2025)}]{park2025maporlmultiagentpostcotrainingcollaborative}
\bibinfo{author}{Park, C.} \emph{et~al.}
\newblock \bibinfo{title}{Maporl: Multi-agent post-co-training for collaborative large language models with reinforcement learning} (\bibinfo{year}{2025}).
\newblock \urlprefix\url{https://arxiv.org/abs/2502.18439}.
\newblock \bibinfo{eprint}{{\href{https://arxiv.org/abs/2502.18439}{{arXiv:2502.18439}}}}.

\bibitem[{Motwani et~al.(2025)}]{motwani2025maltimprovingreasoningmultiagent}
\bibinfo{author}{Motwani, S.~R.} \emph{et~al.}
\newblock \bibinfo{title}{Malt: Improving reasoning with multi-agent llm training} (\bibinfo{year}{2025}).
\newblock \urlprefix\url{https://arxiv.org/abs/2412.01928}.
\newblock \bibinfo{eprint}{{\href{https://arxiv.org/abs/2412.01928}{{arXiv:2412.01928}}}}.

\bibitem[{Liu et~al.(2025)}]{liu2025llmcollaborationmultiagentreinforcement}
\bibinfo{author}{Liu, S.}, \bibinfo{author}{Chen, T.}, \bibinfo{author}{Liang, Z.}, \bibinfo{author}{Lyu, X.} \& \bibinfo{author}{Amato, C.}
\newblock \bibinfo{title}{Llm collaboration with multi-agent reinforcement learning} (\bibinfo{year}{2025}).
\newblock \urlprefix\url{https://arxiv.org/abs/2508.04652}.
\newblock \bibinfo{eprint}{{\href{https://arxiv.org/abs/2508.04652}{{arXiv:2508.04652}}}}.

\bibitem[{Wan et~al.(2025)}]{wan2025remalearningmetathinkllms}
\bibinfo{author}{Wan, Z.} \emph{et~al.}
\newblock \bibinfo{title}{Rema: Learning to meta-think for llms with multi-agent reinforcement learning} (\bibinfo{year}{2025}).
\newblock \urlprefix\url{https://arxiv.org/abs/2503.09501}.
\newblock \bibinfo{eprint}{{\href{https://arxiv.org/abs/2503.09501}{{arXiv:2503.09501}}}}.

\bibitem[{Park et~al.(2026)}]{park2026teamgymgymteamdecisionmaking}
\bibinfo{author}{Park, C.} \emph{et~al.}
\newblock \bibinfo{title}{Teamgym: A gym for team decision-making by ai agents with distributed information} (\bibinfo{year}{2026}).
\newblock \urlprefix\url{https://arxiv.org/abs/2610.37571}.
\newblock \bibinfo{eprint}{{\href{https://arxiv.org/abs/2610.37571}{{arXiv:2610.37571}}}}.

\bibitem[{Bernstein et~al.(2013)}]{bernstein2013complexitydecentralizedcontrolmarkov}
\bibinfo{author}{Bernstein, D.~S.}, \bibinfo{author}{Zilberstein, S.} \& \bibinfo{author}{Immerman, N.}
\newblock \bibinfo{title}{The complexity of decentralized control of markov decision processes} (\bibinfo{year}{2013}).
\newblock \urlprefix\url{https://arxiv.org/abs/1301.3836}.
\newblock \bibinfo{eprint}{{\href{https://arxiv.org/abs/1301.3836}{{arXiv:1301.3836}}}}.

\bibitem[{Shao et~al.(2024)}]{shao2024deepseekmathpushinglimitsmathematical}
\bibinfo{author}{Shao, Z.} \emph{et~al.}
\newblock \bibinfo{title}{Deepseekmath: Pushing the limits of mathematical reasoning in open language models} (\bibinfo{year}{2024}).
\newblock \urlprefix\url{https://arxiv.org/abs/2402.03300}.
\newblock \bibinfo{eprint}{{\href{https://arxiv.org/abs/2402.03300}{{arXiv:2402.03300}}}}.

\bibitem[{Team et~al.(2025)}]{lingteam2025everystep}
\bibinfo{author}{Team, L.} \emph{et~al.}
\newblock \bibinfo{title}{Every step evolves: Scaling reinforcement learning for trillion-scale thinking model} (\bibinfo{year}{2025}).
\newblock \urlprefix\url{https://arxiv.org/abs/2510.18855}.
\newblock \bibinfo{eprint}{{\href{https://arxiv.org/abs/2510.18855}{{arXiv:2510.18855}}}}.

\bibitem[{Marek and Ryabinin(2026)}]{marek2026scorecenteringstabilizesoffpolicy}
\bibinfo{author}{Marek, M.} \& \bibinfo{author}{Ryabinin, M.}
\newblock \bibinfo{title}{Score centering stabilizes off-policy reinforcement learning} (\bibinfo{year}{2026}).
\newblock \urlprefix\url{https://arxiv.org/abs/2609.20807}.
\newblock \bibinfo{eprint}{{\href{https://arxiv.org/abs/2609.20807}{{arXiv:2609.20807}}}}.

\bibitem[{Foerster et~al.(2016)}]{foerster2016learningcommunicatedeepmultiagent}
\bibinfo{author}{Foerster, J.~N.}, \bibinfo{author}{Assael, Y.~M.}, \bibinfo{author}{de~Freitas, N.} \& \bibinfo{author}{Whiteson, S.}
\newblock \bibinfo{title}{Learning to communicate with deep multi-agent reinforcement learning} (\bibinfo{year}{2016}).
\newblock \urlprefix\url{https://arxiv.org/abs/1605.06676}.
\newblock \bibinfo{eprint}{{\href{https://arxiv.org/abs/1605.06676}{{arXiv:1605.06676}}}}.

\bibitem[{Sukhbaatar et~al.(2016)}]{sukhbaatar2016learningmultiagentcommunicationbackpropagation}
\bibinfo{author}{Sukhbaatar, S.}, \bibinfo{author}{Szlam, A.} \& \bibinfo{author}{Fergus, R.}
\newblock \bibinfo{title}{Learning multiagent communication with backpropagation} (\bibinfo{year}{2016}).
\newblock \urlprefix\url{https://arxiv.org/abs/1605.07736}.
\newblock \bibinfo{eprint}{{\href{https://arxiv.org/abs/1605.07736}{{arXiv:1605.07736}}}}.

\bibitem[{Olson(1974)}]{Olson1974-wt}
\bibinfo{author}{Olson, M.}
\newblock \emph{\bibinfo{title}{The logic of collective action}} Harvard Economic Studies (\bibinfo{publisher}{Harvard University Press}, \bibinfo{address}{London, England}, \bibinfo{year}{1974}).

\bibitem[{Leibo et~al.(2017)}]{leibo2017multiagentreinforcementlearningsequential}
\bibinfo{author}{Leibo, J.~Z.}, \bibinfo{author}{Zambaldi, V.}, \bibinfo{author}{Lanctot, M.}, \bibinfo{author}{Marecki, J.} \& \bibinfo{author}{Graepel, T.}
\newblock \bibinfo{title}{Multi-agent reinforcement learning in sequential social dilemmas} (\bibinfo{year}{2017}).
\newblock \urlprefix\url{https://arxiv.org/abs/1702.03037}.
\newblock \bibinfo{eprint}{{\href{https://arxiv.org/abs/1702.03037}{{arXiv:1702.03037}}}}.

\bibitem[{Yang et~al.(2025)}]{yang2025qwen3technicalreport}
\bibinfo{author}{Yang, A.} \emph{et~al.}
\newblock \bibinfo{title}{Qwen3 technical report} (\bibinfo{year}{2025}).
\newblock \urlprefix\url{https://arxiv.org/abs/2505.09388}.
\newblock \bibinfo{eprint}{{\href{https://arxiv.org/abs/2505.09388}{{arXiv:2505.09388}}}}.

\bibitem[{Noukhovitch et~al.(2024)}]{noukhovitch2025asynchronousrlhffasterefficient}
\bibinfo{author}{Noukhovitch, M.} \emph{et~al.}
\newblock \bibinfo{title}{Asynchronous rlhf: Faster and more efficient off-policy rl for language models}.
\newblock \emph{\bibinfo{journal}{ArXiv}} \textbf{\bibinfo{volume}{abs/2410.18252}} (\bibinfo{year}{2024}).
\newblock \urlprefix\url{https://api.semanticscholar.org/CorpusID:273549959}.

\bibitem[{Zelikman et~al.(2022)}]{zelikman2022starbootstrappingreasoningreasoning}
\bibinfo{author}{Zelikman, E.}, \bibinfo{author}{Wu, Y.}, \bibinfo{author}{Mu, J.} \& \bibinfo{author}{Goodman, N.~D.}
\newblock \bibinfo{title}{Star: Bootstrapping reasoning with reasoning} (\bibinfo{year}{2022}).
\newblock \urlprefix\url{https://arxiv.org/abs/2203.14465}.
\newblock \bibinfo{eprint}{{\href{https://arxiv.org/abs/2203.14465}{{arXiv:2203.14465}}}}.

\bibitem[{Gulcehre et~al.(2023)}]{gulcehre2023reinforcedselftrainingrestlanguage}
\bibinfo{author}{Gulcehre, C.} \emph{et~al.}
\newblock \bibinfo{title}{Reinforced self-training (rest) for language modeling} (\bibinfo{year}{2023}).
\newblock \urlprefix\url{https://arxiv.org/abs/2308.08998}.
\newblock \bibinfo{eprint}{{\href{https://arxiv.org/abs/2308.08998}{{arXiv:2308.08998}}}}.

\bibitem[{Lee et~al.(2026)}]{lee2026recursiveselfimprovementmultiagentselfsupervision}
\bibinfo{author}{Lee, H.} \emph{et~al.}
\newblock \bibinfo{title}{Recursive self-improvement through multi-agent self-supervision} (\bibinfo{year}{2026}).
\newblock \urlprefix\url{https://arxiv.org/abs/2610.12176}.
\newblock \bibinfo{eprint}{{\href{https://arxiv.org/abs/2610.12176}{{arXiv:2610.12176}}}}.

\bibitem[{Loshchilov and Hutter(2019)}]{loshchilov2019decoupledweightdecayregularization}
\bibinfo{author}{Loshchilov, I.} \& \bibinfo{author}{Hutter, F.}
\newblock \bibinfo{title}{Decoupled weight decay regularization} (\bibinfo{year}{2019}).
\newblock \urlprefix\url{https://arxiv.org/abs/1711.05101}.
\newblock \bibinfo{eprint}{{\href{https://arxiv.org/abs/1711.05101}{{arXiv:1711.05101}}}}.

\bibitem[{Lowe et~al.(2017)}]{lowe2020multiagentactorcriticmixedcooperativecompetitive}
\bibinfo{author}{Lowe, R.} \emph{et~al.}
\newblock \bibinfo{title}{Multi-agent actor-critic for mixed cooperative-competitive environments} (\bibinfo{year}{2017}).
\newblock \urlprefix\url{https://arxiv.org/abs/1706.02275}.
\newblock \bibinfo{eprint}{{\href{https://arxiv.org/abs/1706.02275}{{arXiv:1706.02275}}}}.

\bibitem[{Lazaridou et~al.(2017)}]{lazaridou2017multiagentcooperationemergencenatural}
\bibinfo{author}{Lazaridou, A.}, \bibinfo{author}{Peysakhovich, A.} \& \bibinfo{author}{Baroni, M.}
\newblock \bibinfo{title}{Multi-agent cooperation and the emergence of (natural) language} (\bibinfo{year}{2017}).
\newblock \urlprefix\url{https://arxiv.org/abs/1612.07182}.
\newblock \bibinfo{eprint}{{\href{https://arxiv.org/abs/1612.07182}{{arXiv:1612.07182}}}}.

\bibitem[{Das et~al.(2020)}]{das2020tarmactargetedmultiagentcommunication}
\bibinfo{author}{Das, A.} \emph{et~al.}
\newblock \bibinfo{title}{Tarmac: Targeted multi-agent communication} (\bibinfo{year}{2020}).
\newblock \urlprefix\url{https://arxiv.org/abs/1810.11187}.
\newblock \bibinfo{eprint}{{\href{https://arxiv.org/abs/1810.11187}{{arXiv:1810.11187}}}}.

\bibitem[{Singh et~al.(2018)}]{singh2018learningcommunicatescalemultiagent}
\bibinfo{author}{Singh, A.}, \bibinfo{author}{Jain, T.} \& \bibinfo{author}{Sukhbaatar, S.}
\newblock \bibinfo{title}{Learning when to communicate at scale in multiagent cooperative and competitive tasks} (\bibinfo{year}{2018}).
\newblock \urlprefix\url{https://arxiv.org/abs/1812.09755}.
\newblock \bibinfo{eprint}{{\href{https://arxiv.org/abs/1812.09755}{{arXiv:1812.09755}}}}.

\bibitem[{Jaques et~al.(2019)}]{jaques2019socialinfluenceintrinsicmotivation}
\bibinfo{author}{Jaques, N.} \emph{et~al.}
\newblock \bibinfo{title}{Social influence as intrinsic motivation for multi-agent deep reinforcement learning} (\bibinfo{year}{2019}).
\newblock \urlprefix\url{https://arxiv.org/abs/1810.08647}.
\newblock \bibinfo{eprint}{{\href{https://arxiv.org/abs/1810.08647}{{arXiv:1810.08647}}}}.

\bibitem[{Wolpert and Tumer(2001)}]{Wolpert2001-hp}
\bibinfo{author}{Wolpert, D.~H.} \& \bibinfo{author}{Tumer, K.}
\newblock \bibinfo{title}{Optimal payoff functions for members of collectives}.
\newblock \emph{\bibinfo{journal}{Adv. Complex Syst.}} \textbf{\bibinfo{volume}{04}}, \bibinfo{pages}{265--279} (\bibinfo{year}{2001}).

\bibitem[{Foerster et~al.(2018)}]{foerster2024counterfactualmultiagentpolicygradients}
\bibinfo{author}{Foerster, J.}, \bibinfo{author}{Farquhar, G.}, \bibinfo{author}{Afouras, T.}, \bibinfo{author}{Nardelli, N.} \& \bibinfo{author}{Whiteson, S.}
\newblock \bibinfo{title}{Counterfactual multi-agent policy gradients} (\bibinfo{year}{2018}).
\newblock \urlprefix\url{https://arxiv.org/abs/1705.08926}.
\newblock \bibinfo{eprint}{{\href{https://arxiv.org/abs/1705.08926}{{arXiv:1705.08926}}}}.

\bibitem[{Rashid et~al.(2018)}]{rashid2018qmixmonotonicvaluefunction}
\bibinfo{author}{Rashid, T.} \emph{et~al.}
\newblock \bibinfo{title}{Qmix: Monotonic value function factorisation for deep multi-agent reinforcement learning} (\bibinfo{year}{2018}).
\newblock \urlprefix\url{https://arxiv.org/abs/1803.11485}.
\newblock \bibinfo{eprint}{{\href{https://arxiv.org/abs/1803.11485}{{arXiv:1803.11485}}}}.

\bibitem[{Yu et~al.(2022)}]{yu2022surprisingeffectivenessppocooperative}
\bibinfo{author}{Yu, C.} \emph{et~al.}
\newblock \bibinfo{title}{The surprising effectiveness of ppo in cooperative, multi-agent games} (\bibinfo{year}{2022}).
\newblock \urlprefix\url{https://arxiv.org/abs/2103.01955}.
\newblock \bibinfo{eprint}{{\href{https://arxiv.org/abs/2103.01955}{{arXiv:2103.01955}}}}.

\bibitem[{Stone et~al.(2010)}]{adHocTeam}
\bibinfo{author}{Stone, P.}, \bibinfo{author}{Kaminka, G.~A.}, \bibinfo{author}{Kraus, S.} \& \bibinfo{author}{Rosenschein, J.~S.} \emph{\bibinfo{title}{Ad hoc autonomous agent teams: collaboration without pre-coordination}}.
\newblock \emph{\bibinfo{booktitle}{Proceedings of the Twenty-Fourth AAAI Conference on Artificial Intelligence}}, AAAI'10, \bibinfo{pages}{1504–1509} (\bibinfo{publisher}{AAAI Press}, \bibinfo{year}{2010}).

\bibitem[{Hu et~al.(2020)}]{hu2021otherplayzeroshotcoordination}
\bibinfo{author}{Hu, H.}, \bibinfo{author}{Lerer, A.}, \bibinfo{author}{Peysakhovich, A.} \& \bibinfo{author}{Foerster, J.}
\newblock \bibinfo{title}{"other-play" for zero-shot coordination} (\bibinfo{year}{2020}).
\newblock \urlprefix\url{https://arxiv.org/abs/2003.02979}.
\newblock \bibinfo{eprint}{{\href{https://arxiv.org/abs/2003.02979}{{arXiv:2003.02979}}}}.

\bibitem[{Strouse et~al.(2022)}]{strouse2022collaboratinghumanshumandata}
\bibinfo{author}{Strouse, D.}, \bibinfo{author}{McKee, K.~R.}, \bibinfo{author}{Botvinick, M.}, \bibinfo{author}{Hughes, E.} \& \bibinfo{author}{Everett, R.}
\newblock \bibinfo{title}{Collaborating with humans without human data} (\bibinfo{year}{2022}).
\newblock \urlprefix\url{https://arxiv.org/abs/2110.08176}.
\newblock \bibinfo{eprint}{{\href{https://arxiv.org/abs/2110.08176}{{arXiv:2110.08176}}}}.

\bibitem[{Jha et~al.(2025)}]{jha2025crossenvironmentcooperationenableszeroshot}
\bibinfo{author}{Jha, K.} \emph{et~al.}
\newblock \bibinfo{title}{Cross-environment cooperation enables zero-shot multi-agent coordination} (\bibinfo{year}{2025}).
\newblock \urlprefix\url{https://arxiv.org/abs/2504.12714}.
\newblock \bibinfo{eprint}{{\href{https://arxiv.org/abs/2504.12714}{{arXiv:2504.12714}}}}.

\bibitem[{Rodionov et~al.(2025)}]{rodionov2025hogwildinferenceparallelllm}
\bibinfo{author}{Rodionov, G.} \emph{et~al.}
\newblock \bibinfo{title}{Hogwild! inference: Parallel llm generation via concurrent attention} (\bibinfo{year}{2025}).
\newblock \urlprefix\url{https://arxiv.org/abs/2504.06261}.
\newblock \bibinfo{eprint}{{\href{https://arxiv.org/abs/2504.06261}{{arXiv:2504.06261}}}}.

\bibitem[{Hsu et~al.(2025)}]{hsu2025groupthinkmultipleconcurrent}
\bibinfo{author}{Hsu, C.-J.} \emph{et~al.}
\newblock \bibinfo{title}{Group think: Multiple concurrent reasoning agents collaborating at token level granularity} (\bibinfo{year}{2025}).
\newblock \urlprefix\url{https://arxiv.org/abs/2505.11107}.
\newblock \bibinfo{eprint}{{\href{https://arxiv.org/abs/2505.11107}{{arXiv:2505.11107}}}}.

\bibitem[{Shinn et~al.(2023)}]{shinn2023reflexionlanguageagentsverbal}
\bibinfo{author}{Shinn, N.} \emph{et~al.}
\newblock \bibinfo{title}{Reflexion: Language agents with verbal reinforcement learning} (\bibinfo{year}{2023}).
\newblock \urlprefix\url{https://arxiv.org/abs/2303.11366}.
\newblock \bibinfo{eprint}{{\href{https://arxiv.org/abs/2303.11366}{{arXiv:2303.11366}}}}.

\bibitem[{Madaan et~al.(2023)}]{madaan2023selfrefineiterativerefinementselffeedback}
\bibinfo{author}{Madaan, A.} \emph{et~al.}
\newblock \bibinfo{title}{Self-refine: Iterative refinement with self-feedback} (\bibinfo{year}{2023}).
\newblock \urlprefix\url{https://arxiv.org/abs/2303.17651}.
\newblock \bibinfo{eprint}{{\href{https://arxiv.org/abs/2303.17651}{{arXiv:2303.17651}}}}.

\bibitem[{Fernando et~al.(2023)}]{fernando2023promptbreederselfreferentialselfimprovementprompt}
\bibinfo{author}{Fernando, C.}, \bibinfo{author}{Banarse, D.}, \bibinfo{author}{Michalewski, H.}, \bibinfo{author}{Osindero, S.} \& \bibinfo{author}{Rocktäschel, T.}
\newblock \bibinfo{title}{Promptbreeder: Self-referential self-improvement via prompt evolution} (\bibinfo{year}{2023}).
\newblock \urlprefix\url{https://arxiv.org/abs/2309.16797}.
\newblock \bibinfo{eprint}{{\href{https://arxiv.org/abs/2309.16797}{{arXiv:2309.16797}}}}.

\bibitem[{Agrawal et~al.(2026)}]{agrawal2026gepareflectivepromptevolution}
\bibinfo{author}{Agrawal, L.~A.} \emph{et~al.}
\newblock \bibinfo{title}{Gepa: Reflective prompt evolution can outperform reinforcement learning} (\bibinfo{year}{2026}).
\newblock \urlprefix\url{https://arxiv.org/abs/2507.19457}.
\newblock \bibinfo{eprint}{{\href{https://arxiv.org/abs/2507.19457}{{arXiv:2507.19457}}}}.

\bibitem[{Zhang et~al.(2026)}]{zhang2026darwingodelmachineopenended}
\bibinfo{author}{Zhang, J.}, \bibinfo{author}{Hu, S.}, \bibinfo{author}{Lu, C.}, \bibinfo{author}{Lange, R.} \& \bibinfo{author}{Clune, J.}
\newblock \bibinfo{title}{Darwin godel machine: Open-ended evolution of self-improving agents} (\bibinfo{year}{2026}).
\newblock \urlprefix\url{https://arxiv.org/abs/2505.22954}.
\newblock \bibinfo{eprint}{{\href{https://arxiv.org/abs/2505.22954}{{arXiv:2505.22954}}}}.

\bibitem[{Zhang et~al.(2026)}]{zhang2026hyperagents}
\bibinfo{author}{Zhang, J.} \emph{et~al.}
\newblock \bibinfo{title}{Hyperagents} (\bibinfo{year}{2026}).
\newblock \urlprefix\url{https://arxiv.org/abs/2603.19461}.
\newblock \bibinfo{eprint}{{\href{https://arxiv.org/abs/2603.19461}{{arXiv:2603.19461}}}}.

\bibitem[{Lin et~al.(2026)}]{lin2026harnessupdatingharnessbenefit}
\bibinfo{author}{Lin, M.} \emph{et~al.}
\newblock \bibinfo{title}{Harness updating is not harness benefit: Disentangling evolution capabilities in self-evolving llm agents} (\bibinfo{year}{2026}).
\newblock \urlprefix\url{https://arxiv.org/abs/2605.30621}.
\newblock \bibinfo{eprint}{{\href{https://arxiv.org/abs/2605.30621}{{arXiv:2605.30621}}}}.

\bibitem[{Liu et~al.(2023)}]{liu2023lostmiddlelanguagemodels}
\bibinfo{author}{Liu, N.~F.} \emph{et~al.}
\newblock \bibinfo{title}{Lost in the middle: How language models use long contexts} (\bibinfo{year}{2023}).
\newblock \urlprefix\url{https://arxiv.org/abs/2307.03172}.
\newblock \bibinfo{eprint}{{\href{https://arxiv.org/abs/2307.03172}{{arXiv:2307.03172}}}}.

\bibitem[{Hong et~al.(2025)}]{trychromaContextRot}
\bibinfo{author}{Hong, K.}, \bibinfo{author}{Troynikov, A.} \& \bibinfo{author}{Huber, J.}
\newblock \bibinfo{title}{{C}ontext {R}ot: {H}ow {I}ncreasing {I}nput {T}okens {I}mpacts {L}{L}{M} {P}erformance --- trychroma.com}.
\newblock \bibinfo{howpublished}{\url{https://www.trychroma.com/research/context-rot\#footnotes}} (\bibinfo{year}{2025}).
\newblock \bibinfo{note}{[Accessed 03-11-2026]}.

\bibitem[{Dang et~al.(2025)}]{dang2025multiagentcollaborationevolvingorchestration}
\bibinfo{author}{Dang, Y.} \emph{et~al.}
\newblock \bibinfo{title}{Multi-agent collaboration via evolving orchestration} (\bibinfo{year}{2025}).
\newblock \urlprefix\url{https://arxiv.org/abs/2505.19591}.
\newblock \bibinfo{eprint}{{\href{https://arxiv.org/abs/2505.19591}{{arXiv:2505.19591}}}}.

\bibitem[{Li et~al.(2026)}]{li2026intheflowagenticoptimizationeffective}
\bibinfo{author}{Li, Z.} \emph{et~al.}
\newblock \bibinfo{title}{In-the-flow agentic system optimization for effective planning and tool use} (\bibinfo{year}{2026}).
\newblock \urlprefix\url{https://arxiv.org/abs/2510.05592}.
\newblock \bibinfo{eprint}{{\href{https://arxiv.org/abs/2510.05592}{{arXiv:2510.05592}}}}.

\bibitem[{Pan et~al.(2025)}]{pan2025learningadaptiveparallelreasoning}
\bibinfo{author}{Pan, J.} \emph{et~al.}
\newblock \bibinfo{title}{Learning adaptive parallel reasoning with language models} (\bibinfo{year}{2025}).
\newblock \urlprefix\url{https://arxiv.org/abs/2504.15466}.
\newblock \bibinfo{eprint}{{\href{https://arxiv.org/abs/2504.15466}{{arXiv:2504.15466}}}}.

\bibitem[{Yang et~al.(2025)}]{yang2025multiverselanguagemodelssecretly}
\bibinfo{author}{Yang, X.}, \bibinfo{author}{An, Y.}, \bibinfo{author}{Liu, H.}, \bibinfo{author}{Chen, T.} \& \bibinfo{author}{Chen, B.}
\newblock \bibinfo{title}{Multiverse: Your language models secretly decide how to parallelize and merge generation} (\bibinfo{year}{2025}).
\newblock \urlprefix\url{https://arxiv.org/abs/2506.09991}.
\newblock \bibinfo{eprint}{{\href{https://arxiv.org/abs/2506.09991}{{arXiv:2506.09991}}}}.

\bibitem[{Subramaniam et~al.(2025)}]{subramaniam2025multiagentfinetuningselfimprovement}
\bibinfo{author}{Subramaniam, V.} \emph{et~al.}
\newblock \bibinfo{title}{Multiagent finetuning: Self improvement with diverse reasoning chains} (\bibinfo{year}{2025}).
\newblock \urlprefix\url{https://arxiv.org/abs/2501.05707}.
\newblock \bibinfo{eprint}{{\href{https://arxiv.org/abs/2501.05707}{{arXiv:2501.05707}}}}.

\bibitem[{Zhou et~al.(2025)}]{zhou2025sweetrltrainingmultiturnllm}
\bibinfo{author}{Zhou, Y.} \emph{et~al.}
\newblock \bibinfo{title}{Sweet-rl: Training multi-turn llm agents on collaborative reasoning tasks} (\bibinfo{year}{2025}).
\newblock \urlprefix\url{https://arxiv.org/abs/2503.15478}.
\newblock \bibinfo{eprint}{{\href{https://arxiv.org/abs/2503.15478}{{arXiv:2503.15478}}}}.

\bibitem[{Liu et~al.(2026)}]{liu2026spiralselfplayzerosumgames}
\bibinfo{author}{Liu, B.} \emph{et~al.}
\newblock \bibinfo{title}{Spiral: Self-play on zero-sum games incentivizes reasoning via multi-agent multi-turn reinforcement learning} (\bibinfo{year}{2026}).
\newblock \urlprefix\url{https://arxiv.org/abs/2506.24119}.
\newblock \bibinfo{eprint}{{\href{https://arxiv.org/abs/2506.24119}{{arXiv:2506.24119}}}}.

\bibitem[{Williams(1992)}]{Williams1992-cw}
\bibinfo{author}{Williams, R.~J.}
\newblock \bibinfo{title}{Simple statistical gradient-following algorithms for connectionist reinforcement learning}.
\newblock \emph{\bibinfo{journal}{Mach. Learn.}} \textbf{\bibinfo{volume}{8}}, \bibinfo{pages}{229--256} (\bibinfo{year}{1992}).

\end{thebibliography}

\clearpage
\appendix
\addtocontents{toc}{\protect\setcounter{tocdepth}{2}}

\renewcommand{\contentsname}{Appendix contents}
\tableofcontents

\section{Extended methods}
\label{app:extended-methods}

\textbf{Models and training.} Information aggregation uses Qwen3-0.6B \citep{yang2025qwen3technicalreport} with thinking on and a per-agent allowance of 8,192 work tokens plus 1,024 for the final answer. Decentralized exploration uses Qwen3-1.7B with thinking on, all reasoning retained in each agent's history, and 8,192 work tokens plus one final token per agent. Both use 200 RL updates. Information aggregation uses 32 problems with eight four-agent teams each, for 256 teams per update; decentralized exploration uses 16 problems with eight teams each, for 128 teams. The matched exploration solo learner uses 32 single-agent episodes per problem, for the same 512 agent trajectories per update as an exploration team learner. Decentralized exploration collection permits 196 teams or 784 solo episodes in flight; rollouts may lag the trained policy by up to four updates, as in asynchronous off-policy RL for LLMs \citep{noukhovitch2025asynchronousrlhffasterefficient}. Information aggregation uses learning rate $5\times10^{-7}$; decentralized exploration uses $2.5\times10^{-7}$. Appendix~\ref{app:details} lists the settings and numerical precision.

\textbf{Demonstrations for decentralized exploration.} So that RL starts from a policy that uses the search tree and message tools validly, we first fine-tune on scripted demonstrations. Four scripted explorers run depth-first search in a depth-2 search tree (121 rooms), announce which branch they are taking at each junction, report exhausted branches, read messages, and back out of branches a teammate reported as empty (Appendix~\ref{app:maze-sft}). The planner communicates verified dead ends and uses only what each agent could have observed or read. The solo learner is fine-tuned on matched demonstrations: the same search trees and random seeds, with four explorers that do not communicate. Each fine-tuning stage runs for 50 updates. RL then uses harder depth-3 search trees. The no-communication and individual-reward exploration controls start from the same fine-tuned team policy; separate social and solo runs start from the base model.

\textbf{Baselines.} Table~\ref{tab:conditions} lists the conditions. The main comparison for decentralized exploration is solo-trained best-of-$N$. It sees the same problems and the same number of agent trajectories per update as the team learner, but each trajectory is a single agent rewarded for its own success, grouped with the other 31 attempts at the same problem. At evaluation we run $N$ copies independently with the same per-agent allowance as the team, and count a success if any copy reaches the exit. The no-communication control keeps the team reward but removes messaging, which separates the value of talking from the value of the team objective. The information aggregation control performs online supervised fine-tuning on successful teams, a training signal related to rejection sampling, expert iteration, and recent multi-agent self-improvement \citep{zelikman2022starbootstrappingreasoningreasoning,gulcehre2023reinforcedselftrainingrestlanguage,lee2026recursiveselfimprovementmultiagentselfsupervision}. It imitates all generated actions in successful teams, including redundant messages, and gives failed teams zero weight. Information aggregation has no no-communication or solo baseline, since no agent can succeed alone.

\textbf{Evaluation.} We evaluate starting policies and final policies after 200 updates on a frozen held-out evaluation dataset of 32 problems per task, disjoint from training (Appendix~\ref{app:data}). For each policy and team size $N\in\{1,2,4,6,8,12,16\}$, we use six evaluation seeds with one team or population attempt per problem per seed, fresh names in every episode, and the training allowance per agent. We report the mean and standard error across the six seed means. Besides solve rate, we report \emph{token efficiency}: solve rate divided by the fraction of the team's allowance spent, $\mathrm{SR}\cdot NL / \bar T$, where $\bar T$ is the mean number of tokens generated by the entire population, failures included. Training traces track reward, success, token use, and communication over the course of training. From every episode we build the directed message graph and measure outgoing-delivery concentration, reciprocity, and room overlap. Tests are paired, two-sided $t$ tests on the six seed means, and $d_z$ is the standardized paired effect. Bonferroni correction is applied within pre-specified families that group contrasts by analysis, task, metric, team size, and comparison type; tests at different team sizes are pointwise. Table~\ref{tab:key-tests} lists the main comparisons and each family's size.

\subsection{Score centering and off-policy training}
\label{sec:score-centering}

Inference and training use separate engines and numerical precisions. For decentralized exploration inference, model weights use blockwise FP8 quantization  and the KV cache uses FP8 E4M3 with computed scales; activations use BF16. Training uses BF16 forward and backward computation with FP32 optimization parameters, AdamW \citep{loshchilov2019decoupledweightdecayregularization} states, and gradient reduction. Information aggregation inference uses BF16 model weights and activations, with FP32 optimization and gradient reduction during training. Furthermore, we run training and inference asynchronously with in-flight updates \citep{noukhovitch2025asynchronousrlhffasterefficient}: inference continues collecting episodes while the trainer updates, with a maximum accepted lag of four updates.

These numerical differences and the lag between collection and training cause the sampling distribution $\mu$ to differ from the trained policy $\pi_\theta$. To account for this mismatch, we use IcePop  \citep{lingteam2025everystep}, which masks tokens with extreme importance ratios. However, masking introduces bias that leads to compounding drift during training\citep{marek2026scorecenteringstabilizesoffpolicy}. To remove this drift from the policy-gradient signal, we apply additive score-centering (SC) correction proposed by \citet{marek2026scorecenteringstabilizesoffpolicy}.

Let $s_\theta(a,c)=\nabla_\theta\log\pi_\theta(a\mid c)$ be the score for token action $a$ and define the masked importance weight
\begin{equation}
w_\theta(a,c)=\mathbf 1\!\left[0.2\leq\frac{\pi_\theta(a\mid c)}{\mu(a\mid c)}\leq5.0\right]\frac{\pi_\theta(a\mid c)}{\mu(a\mid c)}.
\end{equation}
For each generated-token context $c_t$, let $b_\theta(c_t)=\mathbb E_{u\sim\mu(\cdot\mid c_t)}[w_\theta(u,c_t)s_\theta(u,c_t)]$ be the sampler's expected weighted score. The SC corrected gradient direction is
\begin{equation}
g_t^{\rm SC}
 = A_g\bigl[w_\theta(a_t,c_t)s_\theta(a_t,c_t)-b_\theta(c_t)\bigr].
\label{eq:score-centering}
\end{equation}
We estimate $b_\theta$ using independent token draws $u_{t,k}\sim\mu(\cdot\mid c_t)$ with a given context $c_t$, where $K$ is SC's sample-count hyperparameter:
\begin{equation}
\widehat b_\theta(c_t)
 = \frac{1}{K}\sum_{k=1}^{K}
   w_\theta(u_{t,k},c_t)s_\theta(u_{t,k},c_t).
\label{eq:mc-score-centering}
\end{equation}
We use $K=32$ in the decentralized exploration experiments. The draws and their sampler log probabilities are recorded during inference. The correction uses detached importance weights multiplying the score gradients.

Since the smaller information aggregation model is inexpensive to run, we do not use mixed precision in the inference engine and disable SC.
Algorithm~\ref{alg:s3} in Appendix~\ref{app:algorithm} gives the shared collection and team-weighting procedure.

\section{Additional related work}
\label{sec:related}

\textbf{Decentralized coordination.} Decentralized partially observable Markov decision processes formalize cooperative decision-making with separate agent observations and decentralized policies \citep{bernstein2013complexitydecentralizedcontrolmarkov}; multi-agent reinforcement learning studies how such policies can be learned from shared experience and rewards \citep{lowe2020multiagentactorcriticmixedcooperativecompetitive}. Our agents likewise act from separate histories, but their actions include language-model tool calls and voluntary messages, and their shared policy must also work at team sizes not seen in training.

\textbf{Learned communication in multi-agent RL.} Classical multi-agent RL learns communication through differentiable channels \citep{foerster2016learningcommunicatedeepmultiagent,sukhbaatar2016learningmultiagentcommunicationbackpropagation}, discrete referential games \citep{lazaridou2017multiagentcooperationemergencenatural}, targeted messages \citep{das2020tarmactargetedmultiagentcommunication}, learned gates that decide when to speak \citep{singh2018learningcommunicatescalemultiagent}, and rewards for influencing others \citep{jaques2019socialinfluenceintrinsicmotivation}. Credit assignment under a shared reward has been studied through difference rewards \citep{Wolpert2001-hp}, counterfactual baselines \citep{foerster2024counterfactualmultiagentpolicygradients}, value factorization \citep{rashid2018qmixmonotonicvaluefunction}, and centralized critics \citep{lowe2020multiagentactorcriticmixedcooperativecompetitive}, while independent PPO with a shared reward is a strong baseline \citep{yu2022surprisingeffectivenessppocooperative}. Our agents communicate in natural language through tools, and we use comparisons between independent team episodes as the baseline instead of estimating each agent's contribution.

\textbf{Ad hoc teamwork and generalization.} Zero-shot coordination asks agents to work with partners they did not train with \citep{adHocTeam,hu2021otherplayzeroshotcoordination,strouse2022collaboratinghumanshumandata}, and training across many environments yields cooperative norms that transfer to new partners and problems \citep{jha2025crossenvironmentcooperationenableszeroshot}. We ask the analogous question across team sizes.

\textbf{Designed multi-agent systems.} Most LLM teams fix how agents interact. Agents debate in rounds \citep{du2023improvingfactualityreasoninglanguage}, play assigned roles in a software company or a conversation \citep{li2023camelcommunicativeagentsmind,qian2024chatdevcommunicativeagentssoftware,hong2024metagptmetaprogrammingmultiagent}, aggregate each other's samples in layers \citep{wang2024mixtureofagentsenhanceslargelanguage}, or work under an orchestrator that spawns and merges subagents \citep{wu2023autogenenablingnextgenllm,fourney2024magenticonegeneralistmultiagentsolving,anthropicBuiltMultiagent}. Closest in spirit, agents in \citet{park2026scalingdiscoverytesttimecommunication} claim distinct approaches and share reproducible findings at test time. All of these use frozen models. Self-organizing teams can also learn reusable teamwork strategies from prior collaborations, without a fixed workflow or topology, and outperform their strongest member \citep{pappu2026selforganizingagentteamslearn}; these teams are not trained with RL and do not act asynchronously.

\textbf{Parallel reasoning without training.} Parallel instances of one model can attend to each other's partial outputs through a shared cache \citep{rodionov2025hogwildinferenceparallelllm} or interleave threads at the token level \citep{hsu2025groupthinkmultipleconcurrent}, and independent sampling with voting or verification improves steadily with the number of samples \citep{wang2023selfconsistencyimproveschainthought,brown2024largelanguagemonkeysscaling,snell2024scalingllmtesttimecompute}.

\textbf{Searching over structures.} A second line keeps model weights fixed and searches over the structure itself: prompts and edges of an agent graph \citep{zhuge2024languageagentsoptimizablegraphs}, communication topologies proposed by a graph network \citep{zhang2025gdesignerarchitectingmultiagentcommunication}, workflows found by tree search \citep{zhang2025aflowautomatingagenticworkflow}, agent designs written as code \citep{hu2025automateddesignagenticsystems}, or architectures sampled from a supernet \citep{zhang2025multiagentarchitecturesearchagentic}. ReSo assigns agents to subtasks with a learned collaborative reward \citep{zhou2025resorewarddrivenselforganizingllmbased}, and Evolving Agent Teams uses a central supervisor to restructure the team when progress stalls \citep{cao2026evolvingagentteams}. These methods learn which designed structure to use, not how agents should behave inside one.

\textbf{Self-improvement through prompts and harnesses.} Many methods improve the instructions or harness that a fixed model follows, through reflection \citep{shinn2023reflexionlanguageagentsverbal,madaan2023selfrefineiterativerefinementselffeedback}, prompt evolution \citep{fernando2023promptbreederselfreferentialselfimprovementprompt,agrawal2026gepareflectivepromptevolution}, or self-modifying agents \citep{zhang2026darwingodelmachineopenended,zhang2026hyperagents}. Gains from evolved harnesses depend on model size, with mid-sized models benefiting most \citep{lin2026harnessupdatingharnessbenefit}, and populations of self-improving agents that can copy each other become more efficient but not more effective than independent agents at the same cost \citep{jha2026solosociallearningcharacterizing}. Putting collaborative skill in the weights instead leaves the context for the task itself, which matters because performance degrades as context grows \citep{liu2023lostmiddlelanguagemodels,trychromaContextRot}.

\textbf{Training a central controller.} Other work trains weights, but only for a controller that decides the structure. An orchestrator is trained with RL to choose which agent acts next \citep{dang2025multiagentcollaborationevolvingorchestration} or to dispatch parallel subagents while the subagents stay fixed \citep{kimiteam2026kimik25visualagentic}, a planner is trained inside a fixed agent pipeline \citep{li2026intheflowagenticoptimizationeffective}, or one model learns to fork and join its own reasoning threads \citep{pan2025learningadaptiveparallelreasoning,yang2025multiverselanguagemodelssecretly}. The organization is still centralized by design.

\textbf{Training team weights for coordination.} The closest prior approach to ours fine-tunes the model on its own successful multi-agent trajectories. \citet{lee2026recursiveselfimprovementmultiagentselfsupervision} alternate search over multi-agent workflows with supervised fine-tuning on trajectories the model judged successful, and \citet{subramaniam2025multiagentfinetuningselfimprovement} fine-tune separate copies of a model on debate outputs. Our information aggregation control uses verified success as its online fine-tuning signal, with matched collection and update opportunities (Section~\ref{sec:results}). Multi-agent RL for LLMs has trained debate \citep{park2025maporlmultiagentpostcotrainingcollaborative}, generator, verifier and refiner roles \citep{motwani2025maltimprovingreasoningmultiagent}, a meta-thinking agent above a reasoning agent \citep{wan2025remalearningmetathinkllms}, agents that collaborate with a simulated human \citep{zhou2025sweetrltrainingmultiturnllm}, self-play in zero-sum games \citep{liu2026spiralselfplayzerosumgames}, and group decisions in which key facts are split among agents \citep{park2026teamgymgymteamdecisionmaking}. These train a small, fixed number of agents with assigned roles or turn order. MAGRPO \citep{liu2025llmcollaborationmultiagentreinforcement} also shares a group-relative advantage across cooperating agents, but agents act in synchronous turns, each has its own parameters, and the group is formed per turn from joint actions. Our agents share weights, act asynchronously, decide for themselves when and to whom to speak, and are compared as whole episodes, so one policy can be deployed at team sizes it never saw.

\textbf{Avalon-M1.} Concurrent work trains a 120B mixture-of-experts model (12B active) to run agent organizations with one collective advantage \citep{generalreasoning2026avalonm1}. A CEO agent designs the organization, hires up to eight specialists with roles it writes, and can hire and fire them; we impose no structure at all, during training or deployment. Avalon studies how to run an organization; we study how organization emerges from identical agents. Avalon agents exchange messages and shared files and can read any teammate's full trace, so a gain could come from one agent effectively seeing everything; ours share information only through messages they choose to send. Avalon sums rewards over all agent trajectories and penalizes organizations that hire fewer agents than their budget allows; we use the team's best outcome and weight teams equally, so the objective does not favor more agents unless they contribute. Avalon starts RL from a model fine-tuned on GLM-5.2 traces and does not separate the contributions of fine-tuning and RL; it also notes that it does not train a single-agent model on the same environments, and that at matched tokens a single agent with extended thinking recovers about three fifths of its gain. We compare against a solo learner trained with the same RL budget and report fine-tuning-only, RL-only, and ablated controls. Finally, we train with a fixed team size and test generalization to other sizes, and we work with small open models.

\section{Training algorithm and controls}
\label{app:algorithm}

\begin{algorithm}[h]
\caption{\method{}: team-relative RL for decentralized LLM teams}
\label{alg:s3}
\begin{algorithmic}[1]
\Require policy $\pi_\theta$; problems $\mathcal D$; team size $N$; attempts per problem $G$; problems per update $M$; allowance $L$; cost $\lambda$
\For{each update}
  \For{each of $M$ problems $x\sim\mathcal D$} \Comment{episodes run concurrently, subject to the in-flight limit}
    \For{$g=1,\dots,G$}
      \State Sample $N$ distinct names from \texttt{agent\_1..agent\_64}; shuffle each agent's partner list
      \State Open a fresh message board and give each agent its observation and tools
      \State Run agents asynchronously until success, all agents idle or finished, or allowances are spent
      \State Compute $R_g$ (Equation~\ref{eq:reward})
    \EndFor
    \State $A_g\gets (R_g-\bar R)/(\sigma_R+\epsilon)$ \Comment{Equation~\ref{eq:advantage}}
    \State Assign $A_g$ to every generated token of every member of team $g$; mask all received context
  \EndFor
  \State Average token losses within each team, then over teams (Equation~\ref{eq:loss})
  \State Update $\theta$; send weights to the inference server; rollouts may lag by up to four updates
\EndFor
\end{algorithmic}
\end{algorithm}

All $G$ attempts at a problem share a dispatch policy version, and every member of a team receives the same reward; we check both. Complete groups form the $MG$ teams collected for an update. Groups with infrastructure failures are discarded and recollected rather than scored as zero. We use $\epsilon=10^{-8}$. Groups in which all teams receive the same reward have zero advantage; they still count toward the collection budget.

\textbf{\method{}: individual reward.} Each agent's reward is its own outcome: $s_i\in\{0,1\}$ times the same token-cost factor as Equation~\ref{eq:reward}. In information aggregation, $s_i$ is whether agent $i$'s own final list is correct. In decentralized exploration, it is whether agent $i$ reached the exit itself; because the episode ends when anyone exits, at most one agent per team is rewarded. Advantages are normalized over all $GN=32$ agents attempting the problem. Collection, team batching, and equal weighting of teams in the loss are unchanged, so only the credit signal differs.

\textbf{Solo learner.} Each episode has one agent with no messaging tools and the same per-agent allowance as a team member. For each problem we run $GN=32$ episodes and normalize advantages over them, and the loss averages tokens within each episode, then over episodes. Per update it therefore sees the same 16 problems, 512 agent trajectories, and per-agent token allowance as the team learner (Table~\ref{tab:matching}).

\textbf{\method{}: no communication.} The team learner with messaging tools and messaging text removed from the prompt. Agents still form a team, receive the team reward, and share one advantage.

\textbf{\method{}: online SFT.} The information aggregation control trains on the generated tokens of successful teams sampled from its current policy. Its successful-team cross-entropy loss is
\begin{equation}
\begin{aligned}
\mathcal L_{\rm SFT}
  &= -\frac{1}{|\mathcal B|}\sum_{g\in\mathcal B}
     \mathbf 1[\text{team }g\text{ succeeds}] \\
  &\quad\cdot\frac{1}{|\mathcal T_g|}
     \sum_{t\in\mathcal T_g}\log\pi_\theta(a_t\mid c_t).
\end{aligned}
\end{equation}
This is related to REINFORCE \citep{Williams1992-cw} with a binary success weight and no baseline. Failed teams have zero loss weight and count toward the collection budget. It uses no group-relative advantage, importance ratio, trust region, or KL term. Nothing is replayed or refilled, and an update with no successful team skips the optimizer step but still counts toward the collection budget.

\section{Design choices in the team objective}
\label{app:design}

\textbf{One normalization per team.} Treating the members of one team as independent samples would count one outcome $N$ times and compare teammates with each other, although they share the same reward.

\textbf{Equal weight per team.} The loss averages tokens within a team and then averages teams, so teams that generate more tokens, because they talk more or think longer, do not dominate the update.

\textbf{The same credit for every action.} Messages, reads, moves, and private reasoning all receive the team's advantage. We do not estimate each agent's contribution, as difference rewards or counterfactual baselines do \citep{Wolpert2001-hp,foerster2024counterfactualmultiagentpolicygradients}. Comparing independent teams on the same problem serves as the baseline instead.

\textbf{Training targets.} Only tokens an agent sampled are trained. Tool results, notifications, and messages received from teammates are in the agent's context but masked from the loss, so an agent is never trained to reproduce what another agent wrote. The exact sampled tokens and their sampling log probabilities are kept across tool calls and interruptions, so the importance ratio in Equation~\ref{eq:loss} compares the trained policy with the exact distribution that produced each token.

\section{Why coordination becomes harder with team size}
\label{app:coordination}

Equation~\ref{eq:coord} treats transfers as independent, but three further effects grow with $N$. Each agent receives $N-1$ messages, and broadcasts from all agents produce $N(N-1)$ deliveries, so contexts grow while each agent's allowance is fixed. An agent's inbox can be in $2^{N-1}$ states depending on which teammates have reported, so a policy must apply a rule that depends on $N$, such as waiting until it has heard from $N-1$ distinct senders, rather than memorize what to do in each state. Finally, with a shared reward, one agent's actions account for a smaller share of the outcome as $N$ grows, so its training signal becomes noisier. None of these affect independent search, where each agent's behavior can ignore team size.

\section{Harness: tools, asynchrony, and episode phases}
\label{app:harness}

\begin{table}[H]
\centering
\small
\caption{Tools available to each agent. Descriptions are shown to the model as tool schemas.}
\label{tab:tools}
\begin{tabularx}{\linewidth}{lX}
\toprule
Tool & Description given to the model \\
\midrule
\texttt{message(to, text)} & ``Send plans, questions, tentative ideas, dead ends, or checked results to one available teammate or all available teammates. Label uncertainty and verification status. Delivery is immediate. Sending does not read incoming messages or wait for replies; use read\_messages when notified.'' \texttt{to} is a teammate's name or \texttt{all}. \\
\texttt{read\_messages()} & ``Read explicitly received unread peer messages.'' Returns up to 16 unread messages with their senders. \\
\texttt{check\_notifications()} & ``Check unread message counts by sender without reading messages.'' The harness also pushes this tool's output on every arrival. \\
\texttt{move(room)} (decentralized exploration) & ``Walk through a door of your current room into the named adjacent room. Returns what you see there. Your position is private: moving never moves a teammate.'' \\
\texttt{look()} (decentralized exploration) & ``Describe your current room again without moving.'' \\
\bottomrule
\end{tabularx}
\end{table}

\textbf{Generation and notifications.} Each agent generates until it emits a complete tool call, ends its response, or spends its allowance. Generation is requested from the inference server in pieces of at most 1,024 tokens, and an agent's response may continue across pieces up to its remaining allowance. The response length is not otherwise limited. Messages arrive asynchronously at any time. If the recipient is generating, the arrival is recorded immediately and its notification is incorporated after the current request finishes: the sampled prefix is kept unchanged, a notification is appended as a tool output, and the agent continues. An idle recipient is woken by the arrival; there are no synchronized notification rounds. A partial tool call is not executed, and a partial final answer is not graded. A notification looks like
\begin{quote}\small\ttfamily\raggedright
\{"unread\_by\_sender": \{"agent\_17": 1\}, "notice": "You have 1 unread message from agent\_17.\textbackslash nUse read\_messages to read them; this notification does not mark messages as read."\}
\end{quote}
A notification is sent only when the unread counts change, and never includes message contents.

\textbf{Idleness and termination.} An agent that ends a response without a tool call becomes idle. It spends no tokens and resumes when a new message arrives. In decentralized exploration, the episode ends as soon as any agent reaches the exit. In information aggregation, an agent can end its own work at any time by replying with a final answer that begins with \texttt{\#\#\#\#}. Its work phase also ends when its work allowance is spent or its context is full, or when every agent is idle or finished. Each agent that has not answered is then asked once for its final answer, with its own reserved allowance of 1,024 tokens: \emph{``Submit the sorted list of all four agents' numbers now. Reply with one line: \#\#\#\# followed by a JSON array of four integers in ascending order, retaining duplicates. The messaging phase has ended.''} The fixed-team runs reported here never add reminders or prompts to keep working.

\textbf{Messaging service.} Messages go through a message board shared only by the members of one episode. Each team attempt has its own board, so attempts at the same problem cannot see each other. A broadcast is delivered to every teammate.

\textbf{Sampling.} Each agent has its own random stream, so the order in which the server answers requests does not change which tokens a given agent samples.

\section{Prompts}
\label{app:prompts}

Prompts are identical across learners and evaluation conditions, except that messaging text and tools are removed when communication is off. Names, teammate lists, numbers, and room names below are filled in per episode.

\textbf{Information aggregation.} The system prompt has five parts. The first states the task:
\begin{quote}\small
``You are in a team solving a private-information number sorting task. Each of four agents receives one English-written number. You initially know only your own number. Share your actual assignment, explicitly read partners' messages, and associate each received number with its sender so repeated messages do not create duplicates. Collect one number per agent and sort by numeric value, retaining equal values from different agents. Every agent can submit the full list. The team earns reward if at least one final answer is correct; there is no message-count reward. Successful rewards are reduced by at most 0.25, in proportion to the tokens all agents generated divided by the team's total generation allowance. Failures pay no token cost, and a correct answer always earns more than a failure.''
\end{quote}
The last two sentences appear only when the token cost is on, as in all runs reported here; without it, the sentence reads ``there is no message-count or length reward.'' The second part is general guidance on communication:
\begin{quote}\small
``Agents work toward the same team objective; no specialist roles are assigned by the harness. A task may give you private information that partners do not know. Work independently when useful. Communicate when you have information, checked evidence, a counterexample, a correction, or a specific question that may change a teammate's next step. Distinguish verified facts from guesses. Read replies explicitly; sending does not read messages. Teammate names are recipients of message, never function names. Use one recipient or all. Do not count repeated claims as independent verification. If evidence conflicts, compare the underlying reasoning. [\ldots] Submit your own checked answer when asked. Examples are hypothetical and contain no evidence about the current problem.''
\end{quote}
The third part states the shared reward: \emph{``All participating agents share the same reward if at least one agent active at the end submits a correct answer. [\ldots] There is no individual-correctness bonus.''} The fourth gives the agent's name and its teammates in shuffled order. The fifth shows two example tool calls: a broadcast (\emph{``My private number is forty-two (42). Please share your own assignment.''}) and a call to \texttt{read\_messages}. The first user turn is
\begin{quote}\small
``Solve this problem with your team: Four agents each have one private number, written in English. Find all four assigned numbers through communication and return their numeric values in ascending order. Include duplicates once per owning agent. You know only your own assignment; other agents do not know it unless you tell them. Final format: \#\#\#\# followed by a JSON array of four integers. Your private number: seventy-three.''
\end{quote}

\textbf{Decentralized exploration.} The system prompt begins with the shared collaboration prompt:
\begin{quote}\small
``You are a decentralized team working together to solve the SAME complete task as efficiently as possible. Everyone has the full task and a private reasoning history. The task tools specify which artifacts or environment state, if any, are shared. There is no leader or assigned specialist. Any member's success solves the task for the team; optimization follows the task's stated candidate-selection and scoring rule. Choose whether communication will help this task. You may work independently, start useful task work immediately, or propose collaboration through message when it is useful. No opening message, role assignment, negotiation, or consensus is required. Invite complementary approaches, and tell partners which part you are investigating so they can avoid duplicating it. [\ldots] Share partial observations, tentative hypotheses, promising directions, counterexample candidates, failed attempts and dead ends, clearly distinguishing guesses from checked facts. [\ldots] Keep messages concise and avoid repeating identical updates, but do not treat uncertainty as a reason to stay silent. Ordinary assistant prose is private: only calling message sends information. This protocol applies unchanged across tasks.''
\end{quote}
It is followed by the task rules:
\begin{quote}\small
``You are exploring a maze of rooms connected by doors to find the exit. Use native tools: move(room) walks through a door of your current room into the named adjacent room and shows what you see there; look() describes your current room again. Only the doors listed for your current room are reachable in one move; to go back, move to the room you arrived from. Every teammate has the same layout and starts in the same room, but each agent has its own private position. A room whose only door leads back is a dead end. Room names carry no information about where the exit is. Keep track of the rooms you have visited and the doors you have not tried yet, so you do not repeat paths. Reaching the exit solves the task for the whole team and ends the episode. There is no limit on moves or messages; only your token allowance limits how much you explore, so keep each step short. If you end a response without a tool call, you wait until a partner sends you a message.''
\end{quote}
Then come the agent's name, its teammates, and \emph{``Your generation allowance is 8192 tokens; your context window is 32768 tokens, including observations.''} Without communication, the collaboration prompt and teammate list are replaced by \emph{``You explore independently. No messaging or shared notes are available.''} The first user turn gives the task and the initial observation of the start room.

\textbf{Harness rules (both tasks).} Communicating agents receive the notification rules: \emph{``While you are working, we will immediately notify you whenever another agent sends you a message, interrupting your current generation if needed. Notifications show unread-message counts by sender. If you have no unread-message notification, there is no need to check your inbox. Use read\_messages when notified; do not repeatedly poll an empty inbox. You can continue reasoning or doing useful work while waiting for a notification.''} All agents receive the rules for idleness: \emph{``A complete tool call runs immediately and you receive its result. [\ldots] If you end a response without a tool call or final answer, you become idle: no tokens are spent while waiting, and a new peer message wakes you. If everyone is idle or finished, you will be asked for your final answer once. Do not repeat statements merely to keep yourself active.''}

\section{Task generation and data}
\label{app:data}

\textbf{Information aggregation.} Each problem lists private integers drawn uniformly from 1 to 99. Training problems have four numbers, one per agent. Development and evaluation problems have sixteen, of which the first $N$ are used, so the same problem can be evaluated at every team size. Numbers are shown to agents as English words. Final answers are graded by exact match against the sorted list of the participants' numbers.

\textbf{Decentralized exploration.} Search trees are generated level by level. From each junction, $b$ doors lead into corridors of $c$ rooms, and the last room of each corridor is the next junction; after $d$ levels, the corridor ends are dead ends, and one of them is chosen uniformly as the exit. Room labels are random numbers with a fixed number of digits. Each room's door order is shuffled once, when the search tree is generated, and is then identical for every agent and every visit. We do not randomize door order per agent, since that would give the team diverse exploration for free. An observation is a JSON object with the current room, its doors, whether it is the exit, the room the agent arrived from, and a notice for dead ends, for example
\begin{verbatim}
{"room": "R4821",
 "doors": ["R1937", "R7550", "R3104", "R6612"],
 "exit": false, "arrived_from": "R1937"}
\end{verbatim}
A search tree episode succeeds only when an agent's \texttt{move} reaches the exit; claims in text are never graded.

\begin{table}[H]
\centering
\small
\caption{Exploration task settings. All search trees have branching $b=3$, corridor length $c=10$, and no loops.}
\label{tab:mazes}
\begin{tabular}{lcc}
\toprule
& SFT on demonstrations & RL and evaluation \\
\midrule
Depth $d$ & 2 & 3 \\
Rooms & 121 & 391 \\
Dead ends (exit candidates) & 9 & 27 \\
Moves on the shortest path & 20 & 30 \\
Moves for a full depth-first search & 240 & 780 \\
Training / development / evaluation search trees & 1,024 / 32 / 32 & 1,024 / 32 / 32 \\
\bottomrule
\end{tabular}
\end{table}

\textbf{Splits.} Training, development, and held-out evaluation sets are generated from different seeds and fixed before any training. Development sets are used to monitor training. The prespecified checkpoint and final evaluations use the fixed evaluation sets; evaluation outcomes are not used to choose settings.

\section{Training details and budget matching}
\label{app:details}

\begin{table}[H]
\centering
\small
\caption{Training settings.}
\label{tab:hparams}
\begin{tabularx}{\linewidth}{@{}>{\raggedright\arraybackslash}p{0.25\linewidth}XX@{}}
\toprule
Setting & Information aggregation & Decentralized exploration \\
\midrule
Model & Qwen3-0.6B, thinking on & Qwen3-1.7B, thinking on, all reasoning retained \\
Initial policy & Base model & 50 SFT updates on demonstrations (Appendix~\ref{app:maze-sft}) \\
RL updates & 200 & 200 \\
Teams per update & 256 (32 problems $\times$ 8 attempts) & 128 (16 problems $\times$ 8 attempts) \\
Team size in training & 4 & 4 \\
Per-agent allowance & 8,192 work + 1,024 final & 8,192 (+1 unused final token) \\
Generation request size & 1,024 tokens & 1,024 tokens \\
Context length & 40,960 & 32,768 \\
Sampling & temperature 1.0, no top-$p$/top-$k$ & temperature 1.0, no top-$p$/top-$k$ \\
Token cost $\lambda$ & 0.25 & 0.25 \\
Concurrent episodes & 256 & 196 teams / 784 solo episodes \\
Maximum policy lag & 4 updates & 4 updates \\
Optimizer & AdamW, $\beta=(0.9,0.999)$, $\epsilon=10^{-8}$ & same \\
Learning rate & $5\times10^{-7}$, constant, no warm-up & $2.5\times10^{-7}$, constant \\
Weight decay / gradient clipping & 0 / norm 1.0 & same \\
Policy-gradient correction & Probability mask and KL penalty & IcePop and MC32 score centering \\
Score-centering samples $K$ & --- & 32 per generated-token context \\
Precision & BF16 inference, FP32 optimizer and gradient reduction & FP8 weights and KV cache, BF16 activations; FP32 optimizer and reduction \\
Checkpoints & every update; retain evaluation checkpoints & same \\
GPUs per run & 2 trainer + 2 inference & 2 trainer + 6 inference \\
\bottomrule
\end{tabularx}
\end{table}

\begin{table}[H]
\centering
\small
\caption{What each learner sees per update in decentralized exploration. Allowances are per agent.}
\label{tab:matching}
\begin{tabular}{lccccc}
\toprule
Learner & Problems & Episodes & Trajectories & Adv.\ group & Loss unit \\
\midrule
\method{} & 16 & 128 teams of 4 & 512 & 8 teams & team \\
\method{}: no communication & 16 & 128 teams of 4 & 512 & 8 teams & team \\
\method{}: individual reward & 16 & 128 teams of 4 & 512 & 32 agents & team \\
Solo learner & 16 & 512 single agents & 512 & 32 agents & agent \\
\bottomrule
\end{tabular}
\end{table}

\textbf{Asynchronous training.} Rollouts and training run at the same time. The inference server keeps generating new episodes while the trainer updates, and new weights are loaded as they become available. For decentralized exploration the maximum accepted policy lag is four updates; stale rollouts are discarded. The in-flight limits of 196 teams and 784 solo episodes are queue bounds, not batch sizes: each team update uses 128 teams and each matched solo update uses 512 single-agent episodes. Generation uses chunks of up to 1,024 tokens, with message notifications delivered as tool results. Decentralized exploration training uses two H200 GPUs for training and six for inference. The social and solo runs fine-tuned on demonstrations use one trainer and seven inference GPUs for their first six and five optimizer updates, respectively.

\subsection{Training trajectories and communication}
\label{app:training-curves}

\begin{figure}[t]
\centering
\sweepfigure{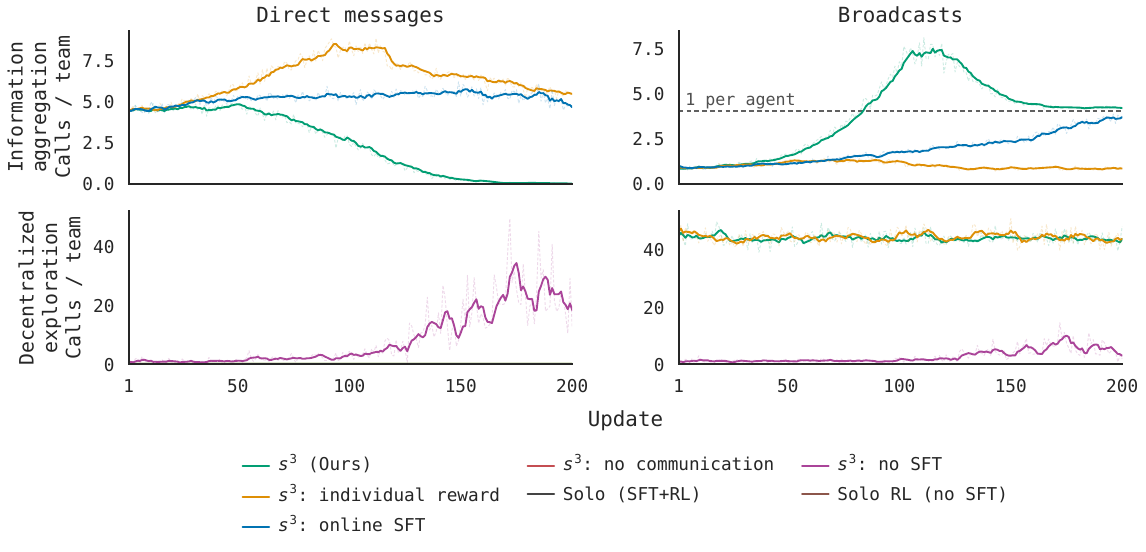}{\linewidth}
\caption{Training communication in information aggregation (top) and decentralized exploration (bottom): mean direct-message calls per team (left) and broadcasts per team (right). In information aggregation, the horizontal dashed line marks four broadcasts per team, one per agent; \method{} approaches this symmetric protocol as direct messages disappear. Light dashed curves show individual updates; solid curves show trailing five-update averages. Learner names follow Table~\ref{tab:conditions}.}
\label{fig:training-communication}
\end{figure}

Table~\ref{tab:training-start-end} reports reward, success, and token efficiency at the start and end of training for every learner. Figure~\ref{fig:training-communication} separates direct messages from broadcasts. In information aggregation, \method{} initially sends repeated broadcasts, then falls to 4.19 broadcasts and 0.03 direct messages per team over updates 181--200, close to one broadcast per agent. A broadcast counts as one message call, irrespective of the number of recipients; solo and communication-disabled learners send no messages.

\begin{figure*}[t]
\centering
\sweepfigure{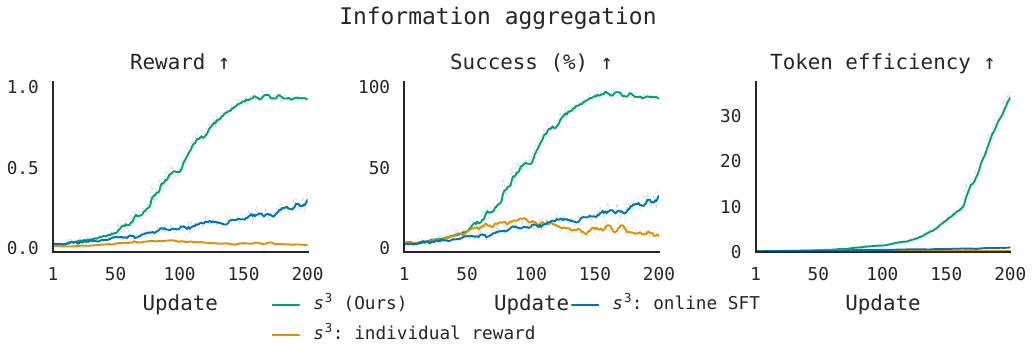}{\linewidth}
\caption{Information aggregation training: reward, solve rate, and token efficiency. Dashed curves show individual updates; solid curves show trailing five-update means. Higher is better. Names follow Table~\ref{tab:conditions}.}
\label{fig:training-curves}
\end{figure*}

\begin{figure*}[t]
\centering
\sweepfigure{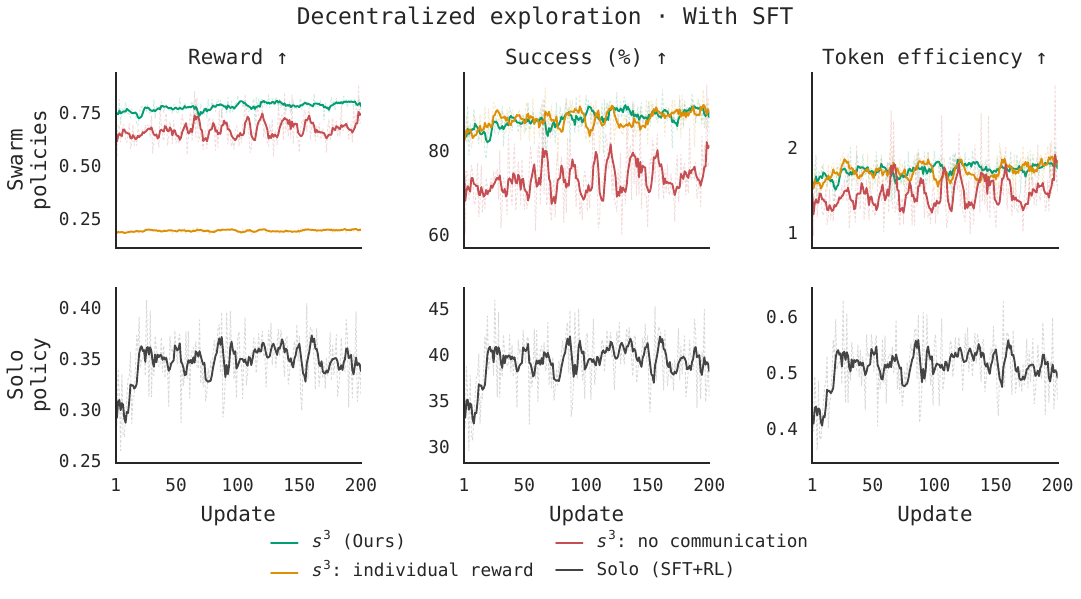}{\linewidth}
\caption{Decentralized exploration training with SFT. Swarm learners appear above and solo learners below. Columns show reward, success, and token efficiency; higher is better. Each panel uses its own vertical range to show learning within the condition. Success is per swarm for swarm learners and per single-agent episode for solo learners. Dashed curves show individual updates; solid curves show trailing five-update means.}
\label{fig:training-exploration-sft}
\end{figure*}

\begin{figure*}[t]
\centering
\sweepfigure{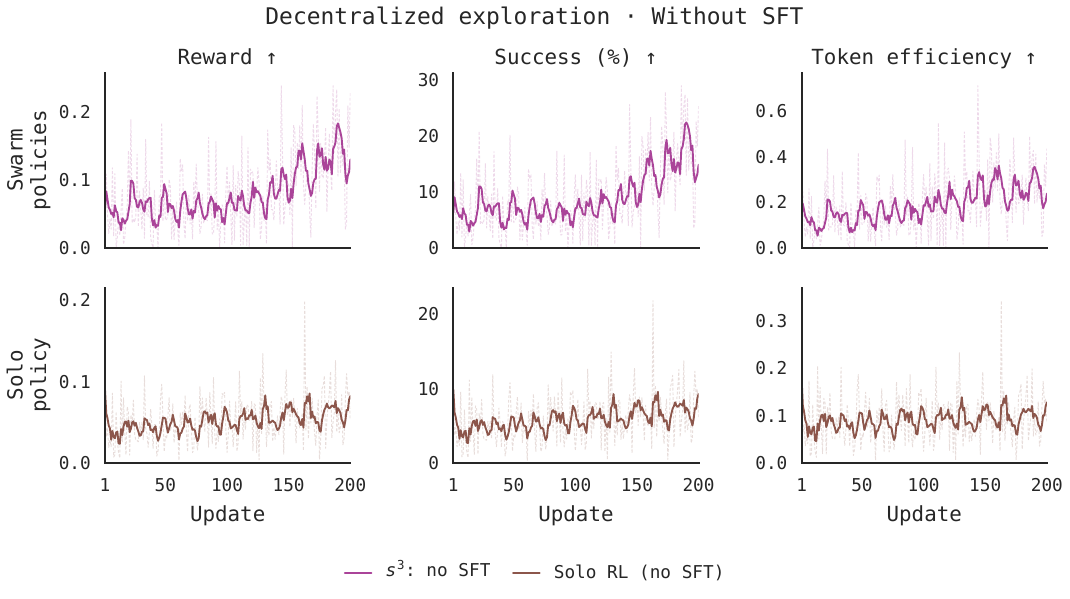}{\linewidth}
\caption{Decentralized exploration training without SFT. Swarm learners appear above and solo learners below. Columns show reward, success, and token efficiency; higher is better. Each panel uses its own vertical range to show learning within the condition. Success is per swarm for swarm learners and per single-agent episode for solo learners. Dashed curves show individual updates; solid curves show trailing five-update means.}
\label{fig:training-exploration-no-sft}
\end{figure*}

\begin{table}[H]
\centering
\small
\setlength{\tabcolsep}{3pt}
\begin{tabularx}{\linewidth}{Xrrrrrrr}
\toprule
& \multicolumn{2}{c}{Reward} & \multicolumn{3}{c}{Success (\%)} & \multicolumn{2}{c}{Efficiency} \\
Learner & Start & End & Start & End & $\Delta$ (pp) & Start & End \\
\midrule
\multicolumn{8}{l}{\textbf{Sorting}} \\
\method{} (Ours) & $0.027\pm0.005$ & $0.926\pm0.002$ & $3.0\pm0.5$ & $93.3\pm0.3$ & +90.3 & $0.09\pm0.02$ & $29.61\pm1.83$ \\
\method{}: individual reward & $0.007\pm0.001$ & $0.018\pm0.001$ & $3.1\pm0.2$ & $8.5\pm0.5$ & +5.4 & $0.09\pm0.01$ & $0.11\pm0.01$ \\
\method{}: online SFT & $0.025\pm0.003$ & $0.262\pm0.012$ & $2.8\pm0.3$ & $28.7\pm1.3$ & +25.9 & $0.08\pm0.01$ & $0.82\pm0.05$ \\
\midrule
\multicolumn{8}{l}{\textbf{Mazes}} \\
\method{} (Ours) & $0.745\pm0.009$ & $0.793\pm0.005$ & $84.0\pm0.9$ & $89.4\pm0.5$ & +5.4 & $1.60\pm0.04$ & $1.78\pm0.02$ \\
\method{}: no communication & $0.643\pm0.004$ & $0.694\pm0.017$ & $70.9\pm0.5$ & $75.8\pm1.8$ & +4.9 & $1.34\pm0.02$ & $1.59\pm0.08$ \\
\method{}: individual reward & $0.187\pm0.001$ & $0.198\pm0.001$ & $84.6\pm0.4$ & $89.3\pm0.5$ & +4.7 & $1.62\pm0.03$ & $1.82\pm0.02$ \\
\method{}: no SFT & $0.051\pm0.003$ & $0.141\pm0.016$ & $5.7\pm0.3$ & $17.1\pm2.0$ & +11.4 & $0.11\pm0.01$ & $0.27\pm0.03$ \\
Solo (SFT+RL) & $0.318\pm0.013$ & $0.342\pm0.003$ & $36.0\pm1.5$ & $38.7\pm0.3$ & +2.7 & $0.46\pm0.02$ & $0.50\pm0.01$ \\
Solo RL (no SFT) & $0.041\pm0.004$ & $0.066\pm0.008$ & $4.6\pm0.4$ & $7.4\pm0.9$ & +2.8 & $0.08\pm0.01$ & $0.10\pm0.01$ \\
\bottomrule
\end{tabularx}
\caption{Training metrics at the start and end of training for every learner. Start and end are means over epochs 1--20 and 181--200. Values are mean $\pm$ SE over four consecutive five-update block means within each window; these SEs describe variation within one training trajectory. $\Delta$ is the change in success rate in percentage points. Higher is better for all three metrics. Reward uses each condition's reward signal; efficiency is solve rate divided by the mean fraction of the token allowance consumed.}
\label{tab:training-start-end}
\end{table}

\begin{figure*}[t]
\centering
\sweepfigure{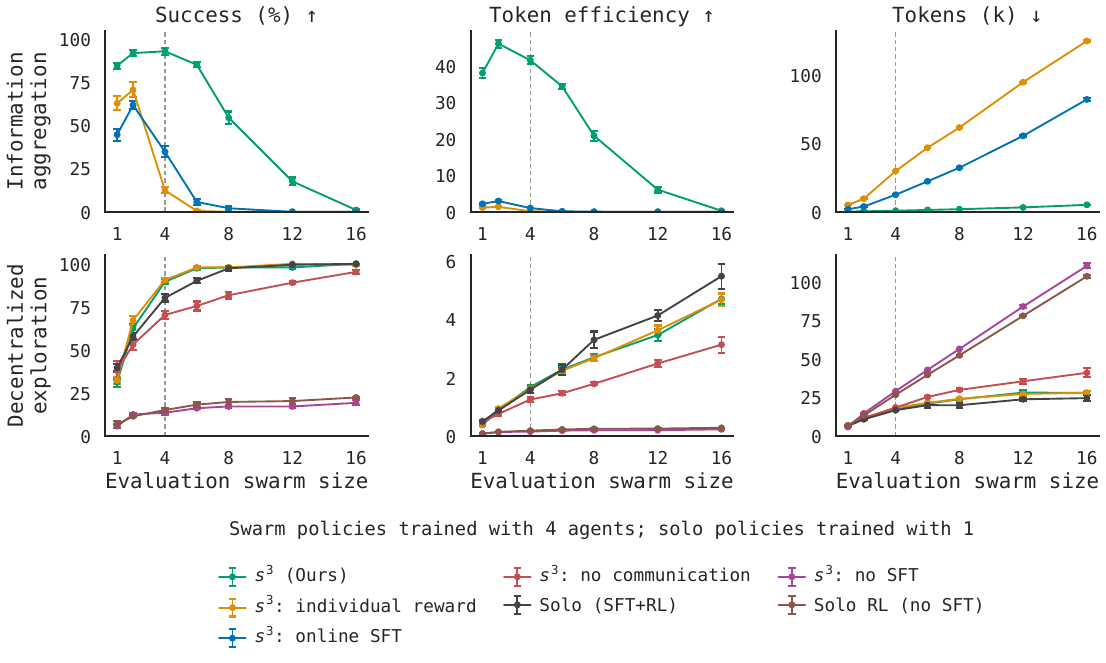}{\linewidth}
\caption{Held-out evaluation across swarm sizes: information aggregation above, decentralized exploration below. All swarm policies were trained with four agents (dashed line); solo policies were trained one agent at a time. Other sizes test generalization without retraining. Columns show success, token efficiency, and total generated tokens per episode, failures included. Token efficiency is solve rate divided by the mean fraction of the population's token allowance used. Higher success and efficiency are better; lower token use is cheaper. Points and bars show mean $\pm$ SE across six evaluation seeds. Learner names follow Table~\ref{tab:conditions}; Appendix Table~\ref{tab:initial-final} compares starting and final policies.}
\label{fig:main}
\end{figure*}
\section{RL gains and the main comparisons}
\label{app:complete-results}

Table~\ref{tab:initial-final} isolates the change produced by training from each learner's actual starting policy. The starting checkpoint is the corresponding SFT policy or base model, evaluated on the same held-out evaluation problems as the final policy. Each value averages 32 problems within seed and then reports mean $\pm$ SE across six seeds.

\begin{figure*}[t]
\centering
\sweepfigure{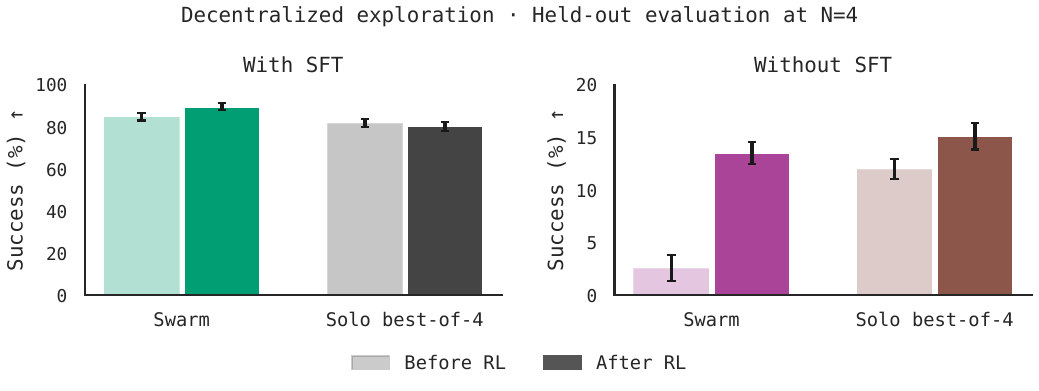}{\linewidth}
\caption{Decentralized exploration before and after RL, with SFT (left) and without SFT (right). Light bars show starting policies; solid bars show final policies. Both panels evaluate four-agent swarms or solo-trained best-of-four on held-out problems. The panels use different success scales to show changes from each starting point. Error bars show SE across six seeds.}
\label{fig:exploration-initial-final}
\end{figure*}
\begin{table}[H]
\centering
\small
\setlength{\tabcolsep}{4pt}
\caption{Starting versus final policy at $N=4$, evaluated on the same 32 held-out evaluation problems and six seeds. Means $\pm$ SE across seed means; gain uses paired seed differences. Starting policies are the corresponding SFT checkpoints or initial base models.}
\label{tab:initial-final}
\begin{tabularx}{\linewidth}{@{}>{\raggedright\arraybackslash}Xrrr@{}}
\toprule
Learner & Start success (\%) & Final success (\%) & Gain (pp) \\
\midrule
Exploration: \method{} (Ours) & $84.9\pm1.9$ & $89.6\pm1.5$ & $4.7\pm1.3$ \\
Exploration: Solo (SFT+RL) & $81.8\pm1.9$ & $80.2\pm2.1$ & $-1.6\pm1.3$ \\
Exploration: \method{}: no communication & $69.8\pm2.6$ & $70.3\pm2.4$ & $0.5\pm2.0$ \\
Exploration: \method{}: individual reward & $84.9\pm1.9$ & $90.6\pm1.4$ & $5.7\pm2.0$ \\
Exploration: \method{}: no SFT & $2.6\pm1.3$ & $13.5\pm1.0$ & $10.9\pm0.7$ \\
Exploration: Solo RL (no SFT) & $12.0\pm1.0$ & $15.1\pm1.3$ & $3.1\pm1.1$ \\
Aggregation: \method{} (Ours) & $1.6\pm1.1$ & $93.2\pm2.0$ & $91.7\pm1.9$ \\
Aggregation: \method{}: individual reward & $1.6\pm1.1$ & $12.5\pm1.8$ & $10.9\pm2.4$ \\
Aggregation: \method{}: online SFT & $1.6\pm1.1$ & $34.9\pm3.2$ & $33.3\pm2.8$ \\
\bottomrule
\end{tabularx}
\end{table}

\begin{figure}[t]
\centering
\sweepfigure{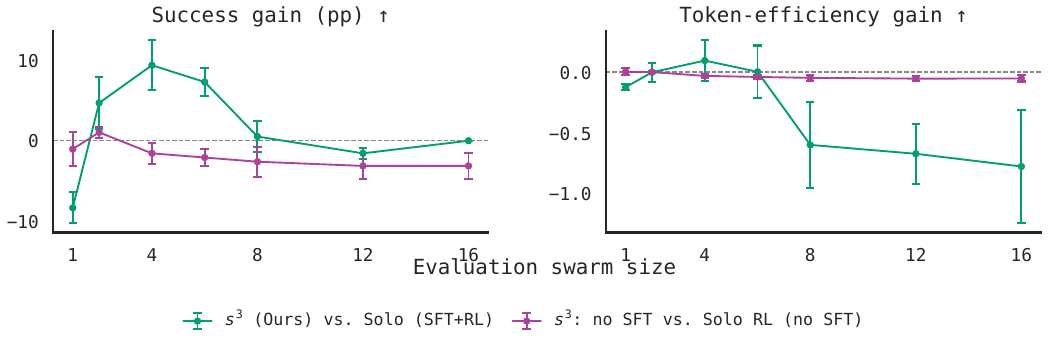}{\linewidth}
\caption{Paired decentralized exploration success gains (left, percentage points) and token-efficiency gains (right) over solo-trained best-of-$N$. \method{} is compared with Solo (SFT+RL), and \method{}: no SFT with Solo RL (no SFT). Positive values favor the team. Points and bars are means $\pm$ SE of six paired seed differences. Table~\ref{tab:paired-maze} gives these main comparisons.}
\label{fig:paired-gains}
\end{figure}

Figure~\ref{fig:paired-gains} and Table~\ref{tab:paired-maze} show where the social team improves on matched solo-trained best-of-$N$. The team fine-tuned on demonstrations is paired with the solo learner fine-tuned on matched demonstrations, and the team trained without SFT with the solo learner trained without SFT. Positive differences favor the team. The largest decentralized exploration success-rate gains occur at four and six agents, before independent search reaches its success ceiling.

\begin{table}[H]
\centering
\small
\setlength{\tabcolsep}{4pt}
\caption{Paired decentralized exploration comparisons against solo-trained best-of-$N$: \method{} versus Solo (SFT+RL), and \method{}: no SFT versus Solo RL (no SFT). Means $\pm$ SE of six within-seed success differences and efficiency differences. Positive differences favor the team.}
\label{tab:paired-maze}
\begin{tabularx}{\linewidth}{@{}>{\raggedright\arraybackslash}Xrrr@{}}
\toprule
Team learner & $N$ & Success gain (pp) & Efficiency gain \\
\midrule
\method{} (Ours) & 1 & $-8.3\pm1.9$ & $-0.13\pm0.03$ \\
\method{} (Ours) & 2 & $4.7\pm3.2$ & $-0.00\pm0.08$ \\
\method{} (Ours) & 4 & $9.4\pm3.1$ & $0.09\pm0.17$ \\
\method{} (Ours) & 6 & $7.3\pm1.7$ & $0.00\pm0.22$ \\
\method{} (Ours) & 8 & $0.5\pm1.9$ & $-0.60\pm0.35$ \\
\method{} (Ours) & 12 & $-1.6\pm0.7$ & $-0.67\pm0.25$ \\
\method{} (Ours) & 16 & $0.0\pm0.0$ & $-0.78\pm0.47$ \\
\method{}: no SFT & 1 & $-1.0\pm2.1$ & $0.00\pm0.03$ \\
\method{}: no SFT & 2 & $1.0\pm0.7$ & $-0.00\pm0.01$ \\
\method{}: no SFT & 4 & $-1.6\pm1.3$ & $-0.03\pm0.02$ \\
\method{}: no SFT & 6 & $-2.1\pm1.0$ & $-0.04\pm0.01$ \\
\method{}: no SFT & 8 & $-2.6\pm1.9$ & $-0.05\pm0.02$ \\
\method{}: no SFT & 12 & $-3.1\pm1.6$ & $-0.05\pm0.02$ \\
\method{}: no SFT & 16 & $-3.1\pm1.6$ & $-0.05\pm0.02$ \\
\bottomrule
\end{tabularx}
\end{table}

\subsection{Tests for the main comparisons}

Tests are two-sided paired $t$ tests on six evaluation seed means, with five degrees of freedom. Cohen's $d_z$ standardizes the paired difference. Bonferroni families separate training-size comparisons, scaling comparisons, ablations, and communication analyses, and are further divided by task, metric, team size, and comparison type. Between-learner contrasts, starting-versus-final contrasts, and within-learner scaling contrasts form separate families. Tests at different team sizes are pointwise. For the main decentralized exploration success comparisons, the family contains two social-versus-solo contrasts (with and without SFT); the decentralized exploration ablation family contains three contrasts. Table~\ref{tab:key-tests} focuses on the main comparisons. Its before/after columns indicate significance at 0.05 before and after correction, with the family size in parentheses. All contrasts in a family count toward its correction, including those omitted from the table.

\begin{table}[p]
\centering
\footnotesize
\setlength{\tabcolsep}{4pt}
\caption{Main comparisons (expl.: decentralized exploration; agg.: information aggregation; train: training-size comparison; scale: larger-population comparison; abl.: ablation; comm.: communication analysis) using two-sided paired $t$ tests on six seed means. Before and after indicate significance at 0.05 before and after Bonferroni correction; parentheses give the family size. Families group contrasts within the same analysis family, task, metric, team size, and comparison type. Tests at different team sizes are pointwise. Effects are first minus second; success differences are percentage points. Cohen's $d_z$ standardizes the paired difference.}
\label{tab:key-tests}
\begin{tabularx}{\linewidth}{@{}p{.65in}>{\raggedright\arraybackslash}Xrp{.58in}rrrr@{}}
\toprule
Family / task & Contrast & $N$ & Metric & Effect $\pm$ SE & $d_z$ & Before & After \\
\midrule
Train / expl. & \method{} vs. Solo (SFT+RL) & 4 & Success & $9.38\pm3.13$ & 1.22 & Yes & No (2) \\
Train / expl. & \method{} vs. Solo (SFT+RL) & 4 & Efficiency & $0.09\pm0.17$ & 0.23 & No & No (2) \\
Scale / expl. & \method{} vs. Solo (SFT+RL) & 6 & Success & $7.29\pm1.74$ & 1.71 & Yes & Yes (2) \\
Scale / expl. & \method{} vs. Solo (SFT+RL) & 6 & Efficiency & $0.00\pm0.22$ & 0.00 & No & No (2) \\
Train / expl. & \method{}: no SFT vs. Solo RL (no SFT) & 4 & Success & $-1.56\pm1.34$ & -0.48 & No & No (2) \\
Abl. / expl. & \method{} vs. \method{}: individual reward & 4 & Success & $-1.04\pm1.55$ & -0.28 & No & No (3) \\
Abl. / expl. & \method{} vs. \method{}: no communication & 4 & Success & $19.27\pm3.56$ & 2.21 & Yes & Yes (3) \\
Abl. / expl. & \method{}: individual reward vs. Solo (SFT+RL) & 4 & Success & $10.42\pm1.92$ & 2.21 & Yes & Yes (3) \\
Abl. / agg. & \method{} vs. \method{}: individual reward & 4 & Success & $80.73\pm2.96$ & 11.15 & Yes & Yes (2) \\
Abl. / agg. & \method{} vs. \method{}: individual reward & 4 & Efficiency & $41.39\pm1.11$ & 15.24 & Yes & Yes (2) \\
Abl. / agg. & \method{} vs. \method{}: online SFT & 4 & Success & $58.33\pm4.17$ & 5.72 & Yes & Yes (2) \\
Abl. / agg. & \method{} vs. \method{}: online SFT & 4 & Efficiency & $40.52\pm1.14$ & 14.53 & Yes & Yes (2) \\
Comm. / expl. & \method{} vs. \method{}: no communication & 4 & Overlap & $-0.19\pm0.01$ & -10.01 & Yes & Yes (3) \\
Comm. / agg. & \method{} vs. \method{}: online SFT & 4 & Reciprocity & $0.46\pm0.01$ & 27.41 & Yes & Yes (2) \\
Scale / agg. & \method{} vs. \method{}: individual reward & 6 & Success & $84.90\pm1.25$ & 27.63 & Yes & Yes (2) \\
Scale / agg. & \method{} vs. \method{}: online SFT & 6 & Success & $79.69\pm2.25$ & 14.48 & Yes & Yes (2) \\
Scale / agg. & \method{} vs. \method{}: individual reward & 8 & Success & $54.69\pm3.59$ & 6.23 & Yes & Yes (2) \\
Scale / agg. & \method{} vs. \method{}: online SFT & 8 & Success & $52.60\pm3.37$ & 6.38 & Yes & Yes (2) \\
Scale / agg. & \method{} vs. \method{}: individual reward & 12 & Success & $17.71\pm2.23$ & 3.24 & Yes & Yes (2) \\
Scale / agg. & \method{} vs. \method{}: online SFT & 12 & Success & $17.71\pm2.23$ & 3.24 & Yes & Yes (2) \\
\bottomrule
\end{tabularx}
\end{table}

\FloatBarrier

\section{Analysis of training trajectories}
\label{app:analysis}

\subsection{Data and method}

\textbf{Sources.} Information aggregation: the per-update team logs of every training team (about 51,000 teams per learner, 155k in total, 1.07M messages), and the full agent traces, with reasoning, at updates 5, 100 and 195 (about 1,000 agents per learner and update). Decentralized exploration: the index of all 531k training episodes for outcomes, and a 20\% sample of the event logs (54k teams) for behavior; the sample's success rates match the full index within 2 points per learner. In decentralized exploration, an episode's update is assigned from its agents' traces.

\textbf{Names.} Names follow Table~\ref{tab:conditions}. \method{} uses shared team reward with communication. Its ablations are \method{}: individual reward (communicating agents rewarded for their own success), \method{}: online SFT (imitation of successful communicating swarms), \method{}: no communication (shared reward without messaging), and \method{}: no SFT (swarm RL from the base model). Solo (SFT+RL) and Solo RL (no SFT) train independent agents, evaluated as best-of-$N$. The trajectory figures use the same names.

\textbf{Error bars.} Bands and bars average each GRPO group (one problem, one policy version) first and then report mean $\pm$ SE across groups, in 10-update bins.

\textbf{Within-group associations.} To relate a team feature to success, we demean success and the feature within each GRPO group (8 information aggregation teams on the same problem at the same policy version) and regress one on the other (linear probability). Slopes are reported in percentage points of success per standard deviation of the feature, with standard errors clustered by group. The 20\% search tree sample has about 1.4 teams per group, so decentralized exploration comparisons use the same search tree and 10-update bin instead. Decentralized exploration teams stop when an agent exits, so search tree message counts are divided by the team's moves, and message content is measured as shares of messages.

\textbf{Communication structure (information aggregation).} Each team falls into one class: at least one agent never sends a message; all agents send and at least one message is direct; all agents only broadcast and some agent broadcasts more than once; or every agent sends exactly one broadcast. Message content is classified by rules that parse English and numeric numbers in the message text and compare them with the team's private numbers.

\textbf{Reasoning.} A turn ends when the agent emits a tool call or ends its response, or when a 1,024-token generation request ends with a teammate's notification pending (Appendix~\ref{app:harness}). We call the last case an interruption. Thought themes are regular expressions over each turn's reasoning text.

\subsection{Information aggregation}

\begin{figure*}[t]
\centering
\includegraphics[width=0.95\textwidth]{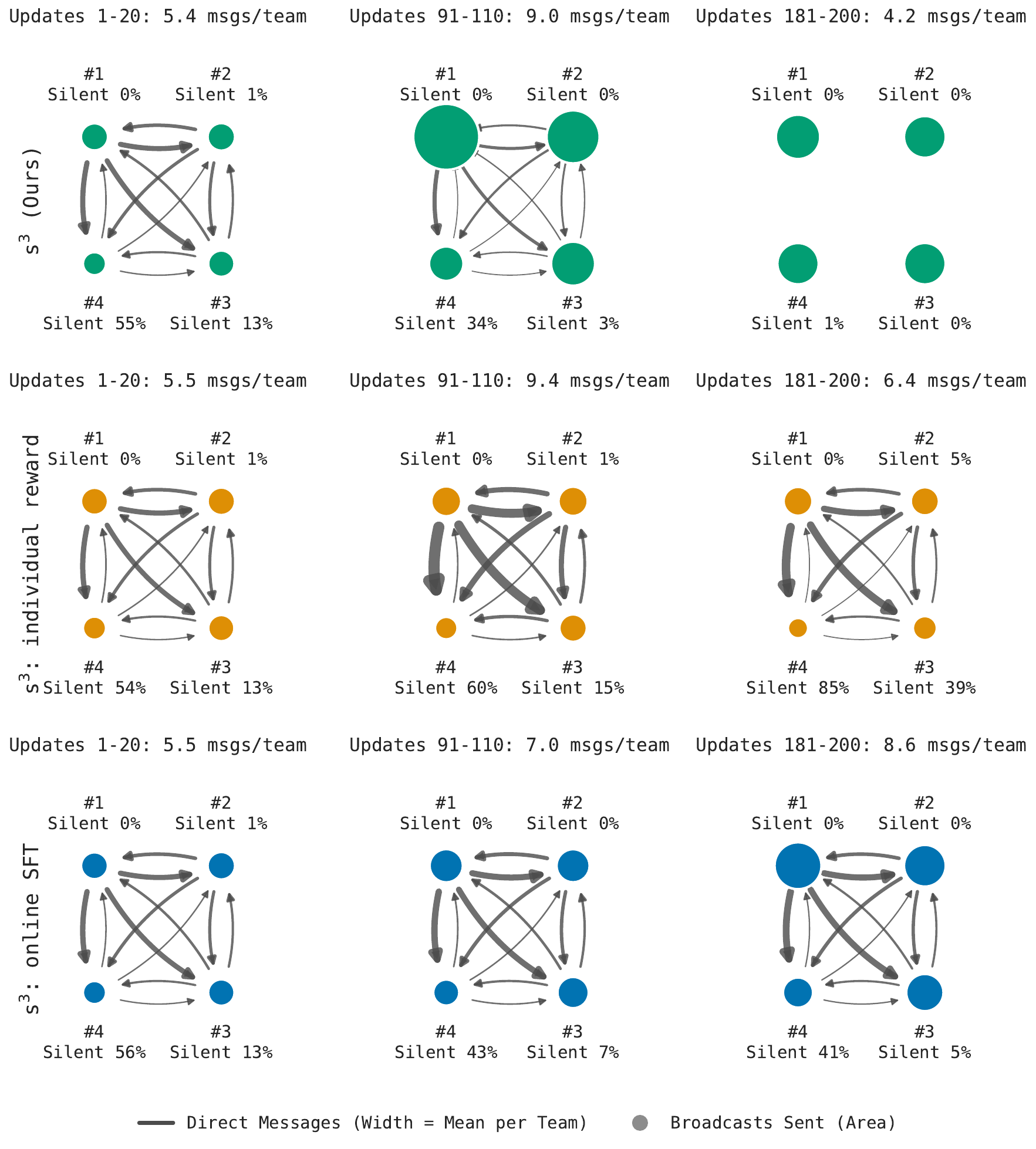}
\caption{Average information aggregation communication graph per learner (rows) and training phase (columns). Within each team, agents are ranked by the number of messages they send (\#1 sends most). Arrow width is the mean number of direct messages per team from one rank to another; node area is the mean number of broadcasts sent, and node colour marks the learner. ``Silent'' is the share of teams in which that agent sends nothing.}
\label{fig:app-graphs}
\end{figure*}

\textbf{Graphs.} Figure~\ref{fig:app-graphs} shows the average graph. All learners start from the same pattern: the most active agent sends about 0.7 direct messages to each teammate, and the least active agent sends nothing in 55\% of teams. Under \method{}, broadcasts take over, and by updates 181--200 each agent broadcasts about once and the inequality of sending across agents (Gini) falls from 0.30 to 0.04. Under individual reward the graph becomes lopsided: the most active agent sends 3.3 direct messages per team, mostly to the two least active agents, which send nothing in 39\% and 85\% of teams, and the Gini rises to 0.45. Online SFT on successes stays mostly direct, with broadcasts rising to 47\% of messages.

\begin{figure*}[t]
\centering
\includegraphics[width=\textwidth]{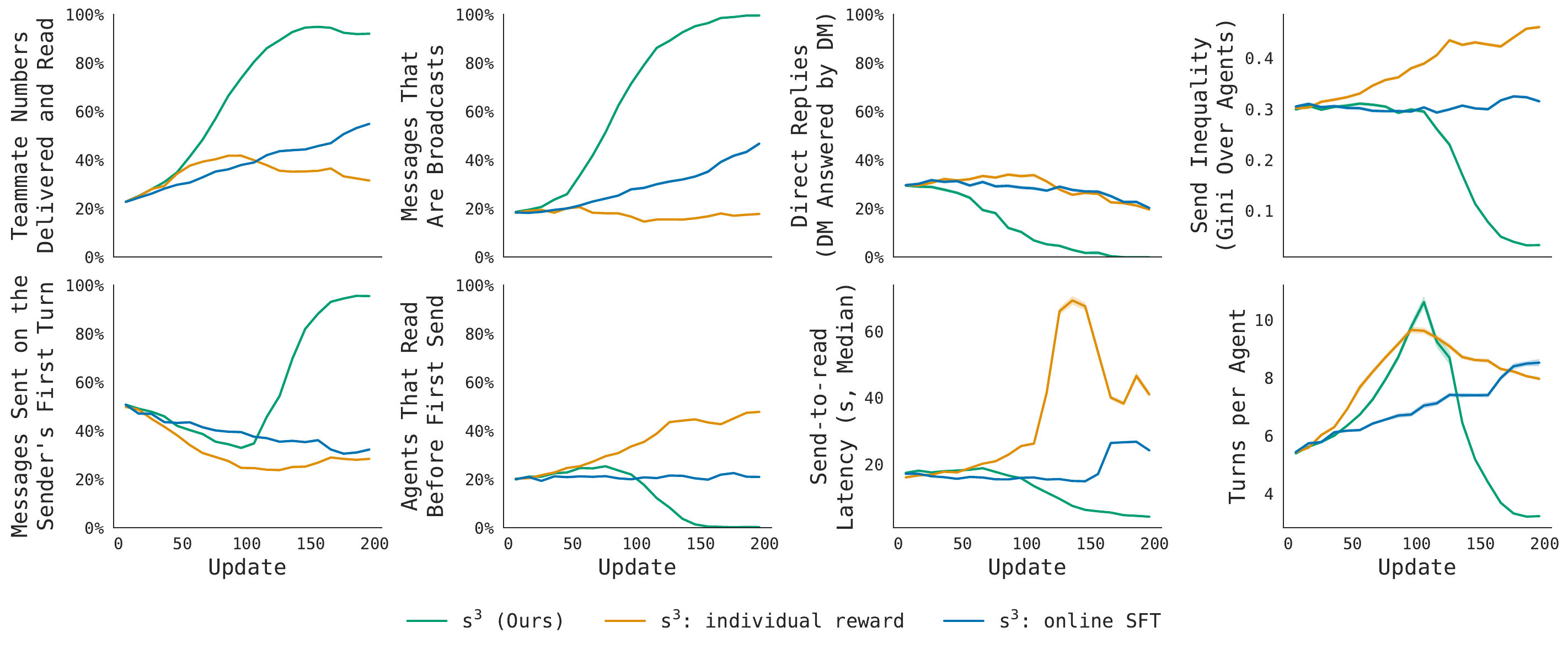}
\caption{Information aggregation communication over training. Top: share of teammate numbers that were both sent and read; share of messages that are broadcasts; share of direct-message pairs answered by a direct message in the other direction; inequality of messages sent across agents. Bottom: share of messages sent on the sender's first turn; share of agents that read their inbox before their first send; median time from send to read; turns per agent. Mean $\pm$ SE across GRPO groups.}
\label{fig:app-cadence}
\end{figure*}

\textbf{Timing.} Under \method{}, messages sent on the sender's first turn dip from 50\% to 33\% around update 100 and then reach 95\%; agents that read before their first send fall from 20\% to 0.4\%; the median time from send to read falls from 17\,s to 5\,s; and turns per agent peak at 10.6 during the repeated-broadcast phase before falling to 3.3 (Figure~\ref{fig:app-cadence}). Under individual reward, agents that read first rise to 48\%, first-turn messages fall to 28\%, and the median time from send to read peaks at 69\,s near update 135, as messages wait for agents that are themselves waiting.

\begin{figure*}[t]
\centering
\includegraphics[width=\textwidth]{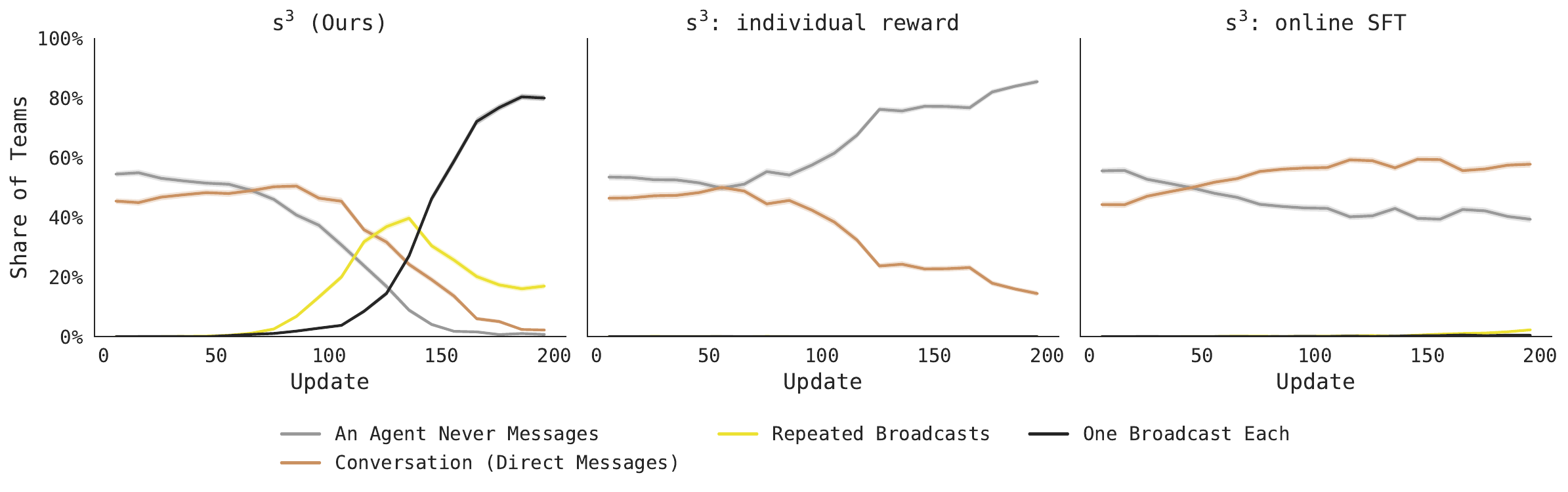}
\caption{Share of information aggregation teams in each communication structure over training, for each learner. Mean $\pm$ SE across GRPO groups.}
\label{fig:app-structure}
\end{figure*}

\begin{table}[H]
\centering
\small
\caption{Information aggregation success by communication structure, relative to conversational teams in the same GRPO group (percentage points, SE clustered by group). The last column gives each structure's share of \method{} teams in updates 101--200.}
\label{tab:app-structure}
\begin{tabular}{lccc}
\toprule
& \multicolumn{2}{c}{\method{}} & \\
Structure & 1--100 & 101--200 & Share \\
\midrule
An agent never messages & $-12.9\pm0.5$ & $-25.9\pm1.3$ & 8\% \\
Conversation (reference) & 0 & 0 & 18\% \\
Repeated broadcasts & $+24.7\pm2.0$ & $+4.4\pm0.7$ & 26\% \\
One broadcast each & $+30.0\pm3.2$ & $+6.9\pm0.7$ & 48\% \\
\bottomrule
\end{tabular}
\end{table}

\textbf{Structure and success.} Figure~\ref{fig:app-structure} shows the structure shares for all learners and Table~\ref{tab:app-structure} their within-group success. Individual reward and online SFT keep the starting mix. In the few teams where they produce a broadcast structure, it also succeeds more often: repeated-broadcast teams under individual reward succeed $74\pm11$ points more often than conversational teams in the same group over updates 101--200, and one-broadcast teams under online SFT $58\pm5$ points more often.

\begin{figure*}[t]
\centering
\includegraphics[width=\textwidth]{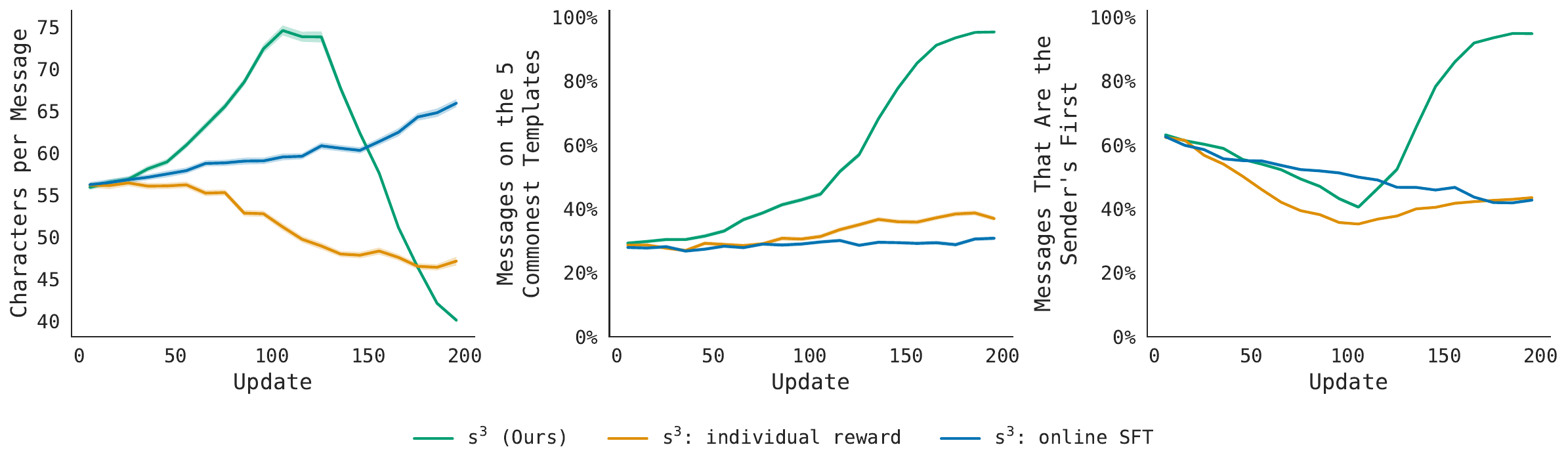}
\caption{Information aggregation message language over training: characters per message; share of messages matching the five most common templates of that learner and 10-update bin (numbers and names replaced by placeholders); share of messages that are their sender's first. Mean $\pm$ SE across GRPO groups.}
\label{fig:app-language}
\end{figure*}

\textbf{Language.} Under \method{}, messages lengthen from 56 to 74 characters during the repeated-broadcast phase, when they carry a number and a request (``My private number is X. Please share your own assignment.''), and shrink to 41 characters at the end; the five most common templates cover 29\% of messages at the start and 95\% at the end (Figure~\ref{fig:app-language}). The example message in the prompt (Appendix~\ref{app:prompts}) leaves traces: ``Please share your own assignment'' appears in 23\% of late \method{} messages, and ``forty-two'' accounts for 25\% of misstated numbers across learners, against 4\% for the next most common value.

\begin{table*}[t]
\centering
\small
\caption{Message content (share of messages) and failure causes (share of failed teams, first matching cause). ``Own number stated as recipient's'' copies the agent's own prompt line (``Your private number is X.'') into a message.}
\label{tab:app-content}
\begin{tabularx}{\linewidth}{@{}>{\raggedright\arraybackslash}X*{4}{>{\centering\arraybackslash}p{1.75cm}}@{}}
\toprule
& Updates 1--20 & \method{}, 101--200 & \method{}: individual reward, 101--200 & \method{}: online SFT, 101--200 \\
\midrule
\multicolumn{5}{l}{\emph{Message content}} \\
Own number (``My private number is forty-two.'') & 71\% & 90\% & 64\% & 74\% \\
Request only (``What is your private number?'') & 21\% & 3\% & 26\% & 16\% \\
Own number stated as recipient's & 4\% & 1\% & 6\% & 3.5\% \\
Misstated own number & 1.6\% & 1.2\% & 1.7\% & 1.9\% \\
\midrule
\multicolumn{5}{l}{\emph{Failure cause}} \\
A number withheld: an agent never messaged & 53\% & 13\% & 74\% & 38\% \\
A number withheld: the agent only asked for others' & 17\% & 1.5\% & 9\% & 15\% \\
A number withheld: misstated or garbled & 8\% & 2\% & 6\% & 9\% \\
Sent but never read & 10\% & 0.3\% & 1.4\% & 8\% \\
Had all numbers, answer contains a wrong number & 6\% & 71\% & 6\% & 22\% \\
Had all numbers, number dropped or duplicated, or list unsorted & 3\% & 12\% & 1.5\% & 4.5\% \\
\bottomrule
\end{tabularx}
\end{table*}

\textbf{Content and failures.} Table~\ref{tab:app-content} classifies messages and failed teams. Early failures are failures to deliver information, mostly because an agent never sends. \method{} removes these, and its remaining failures are transcription errors by agents that had every number. Under individual reward, three quarters of failures have an agent that never messaged.

\begin{figure*}[t]
\centering
\includegraphics[width=\textwidth]{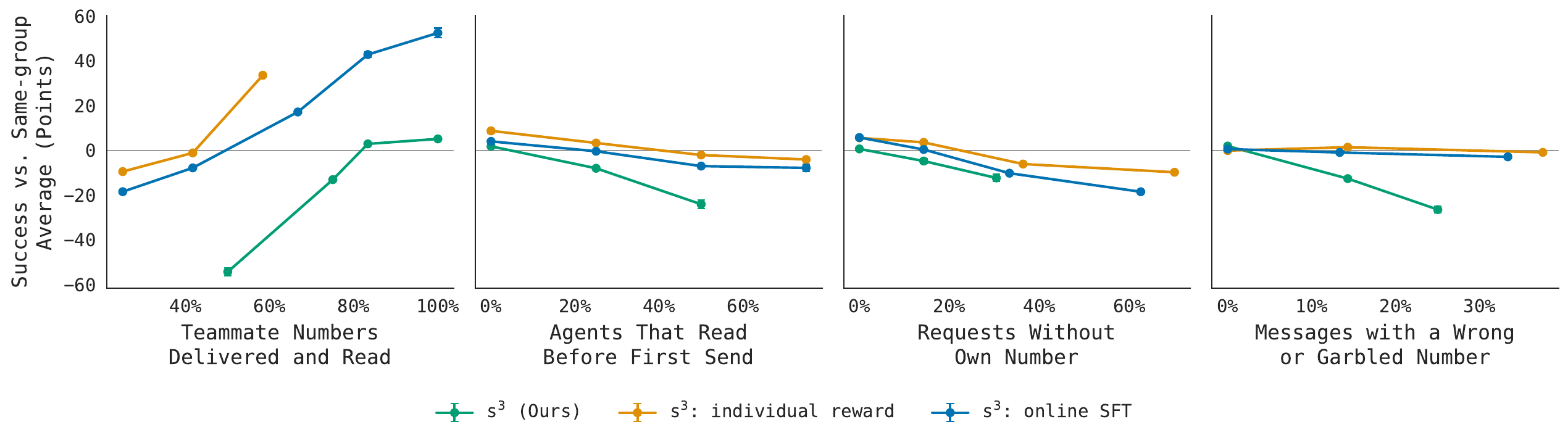}
\caption{Information aggregation success by team feature, updates 101--200. The y-axis is a team's success minus the mean success of its GRPO group (same problem and policy version), in percentage points. Teams are binned by each feature, with points at each bin's median; bins with fewer than 100 teams are omitted. Mean $\pm$ SE across teams. Delivering and reading every teammate's number goes with success under every learner; reading before sending, asking for numbers without giving one's own, and wrong or garbled numbers go with failure. The text gives per-SD slopes for further features.}
\label{fig:app-assoc}
\end{figure*}

\textbf{What goes with success.} Figure~\ref{fig:app-assoc} reports within-group associations. The share of teammate numbers that are sent and read is the strongest ($+12.7$, $+14.5$ and $+20.6$ points per SD under \method{}, individual reward and online SFT). Reading before the first send goes with lower success under every learner ($-7.6$, $-4.2$, $-4.4$), including among teams in which some agent had every number. Asking for a number without giving one's own goes with lower success ($-3.0$, $-7.3$, $-9.3$), and so do direct replies, a direct message answered by one in the other direction ($-1.6$, $-0.5$, $-2.7$). Some associations change sign with the protocol: longer messages and relaying a teammate's number go with lower success under \method{} ($-2.2$ and $-2.1$), whose converged message is short, and with higher success under individual reward ($+4.4$, $+3.3$) and online SFT ($+4.9$, $+4.0$), where a longer message usually carries more numbers. Under \method{}, a wrong or garbled number goes with $-7.3$ points.

\textbf{Silence pays the silent agent.} Within a group over updates 151--200 of individual reward, an agent that never sends is correct 5.6\% of the time, against 1.8\% for agents that send ($+4.3\pm0.2$ points within group), while its team succeeds $6.1\pm0.2$ points less often. The same pattern holds at updates 1--50 for every learner ($+2.4$ to $+3.2$ points for the agent, $-3.5$ to $-4.2$ for the team). Under the team reward, only the team's loss counts.

\begin{figure*}[t]
\centering
\includegraphics[width=\textwidth]{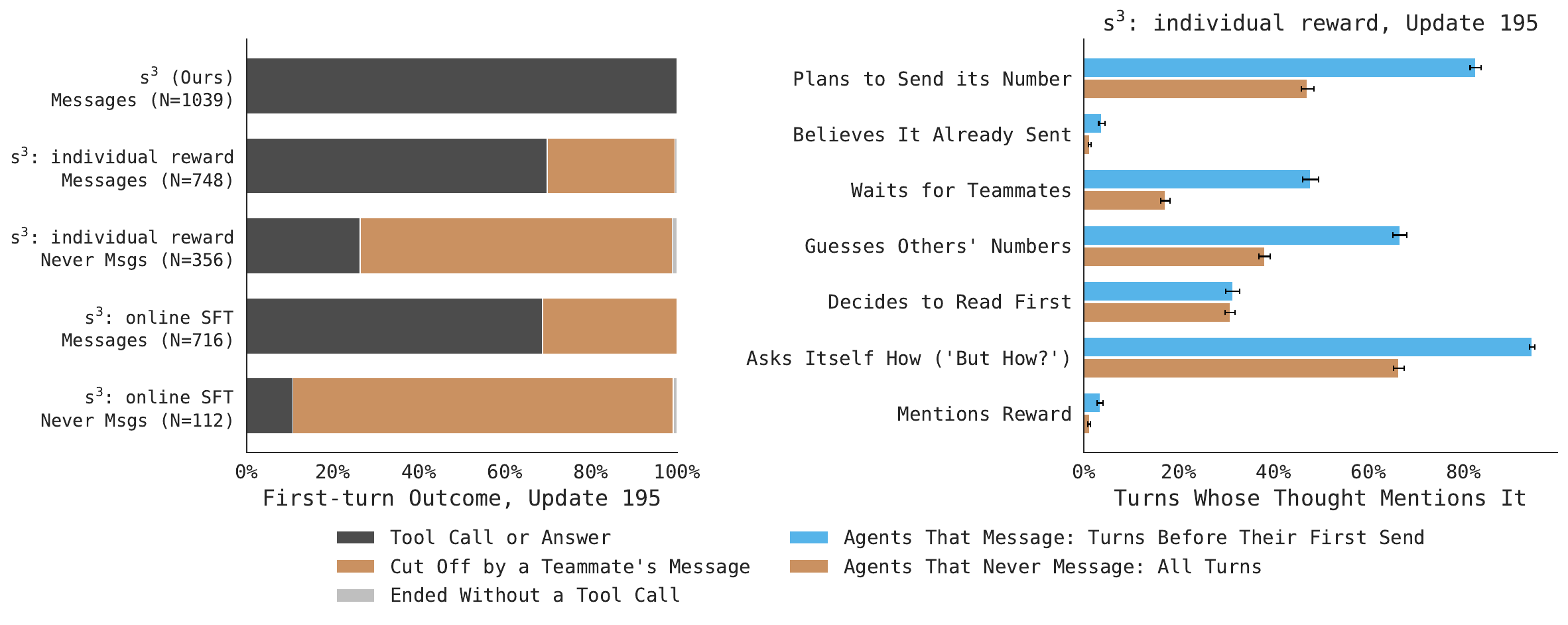}
\caption{Information aggregation reasoning at update 195. Left: how each agent's first turn ended, for agents that message and agents that never message (groups with at least 20 agents). Right: share of turns whose thought matches each theme under individual reward, for agents that message (turns before their first send) and agents that never message (all turns), mean $\pm$ SE across agents.}
\label{fig:app-reasoning}
\end{figure*}

\textbf{Why agents never message.} At update 5, the median first thought runs 700--880 words, about one generation request, and 37--42\% of agents whose first turn is interrupted by a teammate's notification never message, against 11--13\% of the others. At update 195 under individual reward the gap is 54\% against 16\% (Figure~\ref{fig:app-reasoning}). An interrupted agent reads the message, and its later turns concern the inbox: never-messaging agents at update 195 take 6.9 turns on average, 63\% ending in a tool call, mostly reads and the final answer. Their thoughts plan to send their own number in 47\% of turns (82\% for agents that message, before their first send), mention reward or credit in 1\%, and believe they have already sent their number in 1\%. Their final answers fill missing numbers with guesses. Under \method{}, the median first thought falls to 65 words by update 195, and every agent sends before a notification can arrive.

\textbf{Problem features.} Information aggregation problems differ only in their numbers. Over updates 101--200, \method{} succeeds on 86\% of problems with a repeated number and 89--90\% of the rest, regardless of the smallest gap between numbers, and sends 5.7--5.9 messages per team in every gap range. Under individual reward, agents holding the smallest to the largest number send 1.85--1.89 messages each.

\subsection{Decentralized exploration}

\begin{figure*}[t]
\centering
\includegraphics[width=\textwidth]{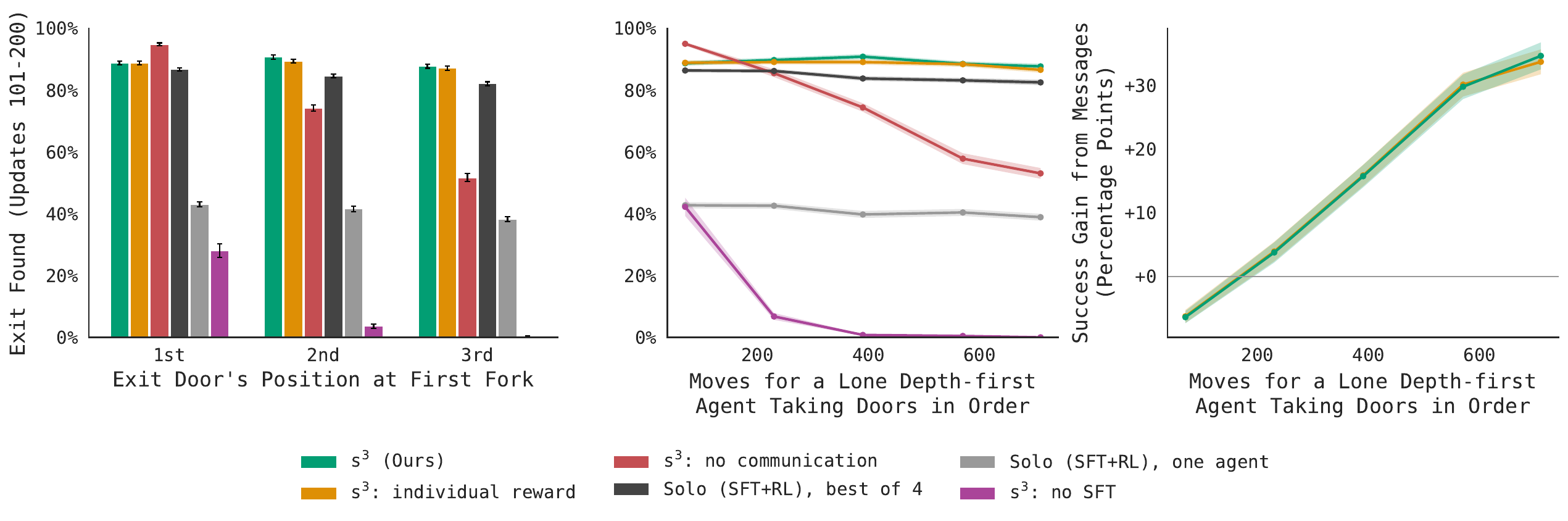}
\caption{Decentralized exploration success by problem, updates 101--200. Left: by the listed position of the exit's door at the first fork. Middle: by the number of moves a lone depth-first agent, taking doors in listed order, needs to reach the exit (quintiles). Right: gain from messages, the success of \method{}: individual reward and of \method{} minus that of \method{}: no communication. Mean $\pm$ SE across search trees; every episode whose GRPO group maps to a search tree (78\% of episodes).}
\label{fig:app-maze-problems}
\end{figure*}

\textbf{Which search trees need messages.} All training search trees have the same shape, so they differ only in the order in which doors are listed. Figure~\ref{fig:app-maze-problems} relates success to that order. Without communication, success falls from 95\% to 53\% across quintiles of the moves a lone depth-first agent needs (median 70 to 710 moves), and the gain from messages rises from $-6$ to $+34$ points. The per-search tree gain correlates 0.67 with the exit's door position at the first fork, $-0.05$ and $0.05$ at the second and third forks, and 0.63 with the depth-first cost; gains measured on updates 101--150 and 151--200 correlate 0.46 across search trees. Communicating teams and solo-trained best-of-four stay within 4 and 5 points across search trees: both spread over the first fork's doors, the teams through claims and the independent agents through sampling (36\%, 32\% and 32\% of first moves). Without communication, success on search trees with the exit behind the third door rises only from 45\% to 52\% over training. Trained from the base model, teams take the first door on 99.9\% of first moves and solve 28\% of first-door search trees against 0.3\% of third-door search trees.

\begin{figure*}[t]
\centering
\includegraphics[width=\textwidth]{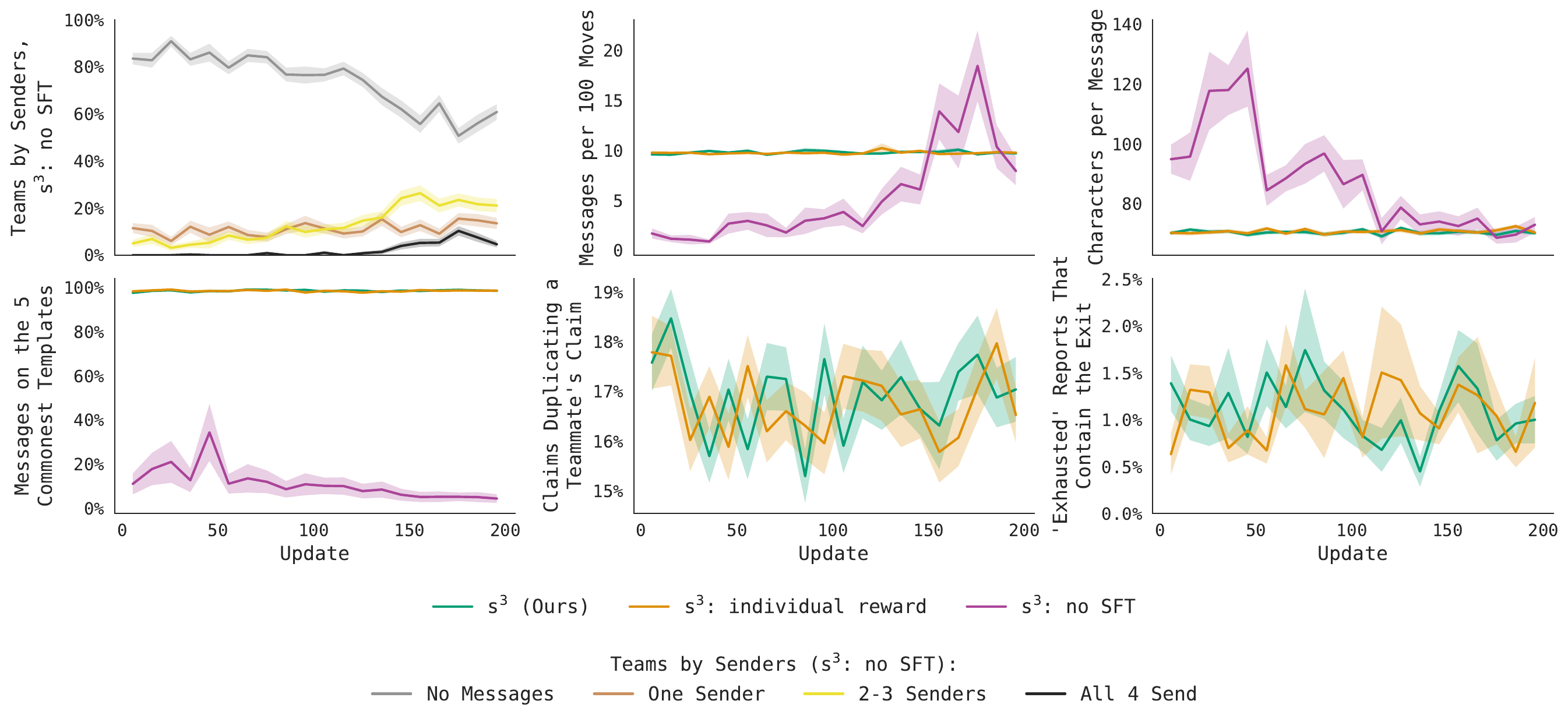}
\caption{Decentralized exploration communication over training. Top left: teams trained from the base model, by the number of agents that send. Other panels: messages per 100 moves; characters per message; share of messages matching the five most common templates; share of branch claims that duplicate a teammate's claim; share of forward ``exhausted'' reports whose branch contains the exit. Mean $\pm$ SE across GRPO groups.}
\label{fig:app-maze-comm}
\end{figure*}

\textbf{Communication over training.} Teams fine-tuned on demonstrations keep the demonstrations' two message templates, a claim (``I'm checking the branch from R9639 through R4496.'') and a report (``Branch from R4496 through R1509 is exhausted; no exit.\ldots''), and stay constant on every measure in Figure~\ref{fig:app-maze-comm}. Every agent's first action is a claim, 99.9\% of messages are broadcasts, and 98\% are read. Because all four agents claim before reading, 40\% of the first four claims duplicate a teammate's; later claims duplicate 14\% of the time. After reading a teammate's claim at the same fork, agents choose a different door 58\% of the time. Trained from the base model, teams with no messages fall from 84\% to 59\%, messages per 100 moves rise from 1.5 to 9.8, the rate of the fine-tuned teams, and messages shorten from 97 to 74 characters, in free text outside the claim and report templates. Early base-model messages are often fabricated (``The exit is in R\ldots. All doors have been checked''); late ones ask for help (``Let's work together to find the exit.'').

\textbf{What changes during decentralized exploration RL.} Fine-tuned \method{} teams improve from 83\% success in updates 1--20 to 89\% in updates 101--200 (all training episodes). Within the same search tree, from updates 1--50 to 151--200 in the 20\% sample, success rises $2.6\pm2.0$ points, moves per team fall by $22\pm17$ (469 to 457), tokens per team by $835\pm623$, the share of teams in which some agent read a false ``exhausted'' report falls from 12.0\% to 10.0\% ($-2.2\pm1.9$ points), and duplicate claims stay at 16--17\% of claims. Individual-reward teams do not improve over the same updates ($-0.7\pm2.2$ points within search tree).

\textbf{Reports.} The sender of a claim enters the claimed branch 98\% of the time. Reports sent after the sender visited every room of the branch (82\% of forward reports, 122k reports) are never false. When the sender saw at most half of the branch, the reported branch contains the exit 13\% of the time, and 65\% of these false reports still state that the sender checked every onward door. In 32\% of failed \method{} teams some agent had read a false report about the exit's branch, and teams with such a report succeed 61\% of the time against 89\% without. Without communication, 44\% of failed teams stop short of 30\% of the exit path, against 5.5--8\% with messages.

\begin{figure}[t]
\centering
\includegraphics[width=\linewidth]{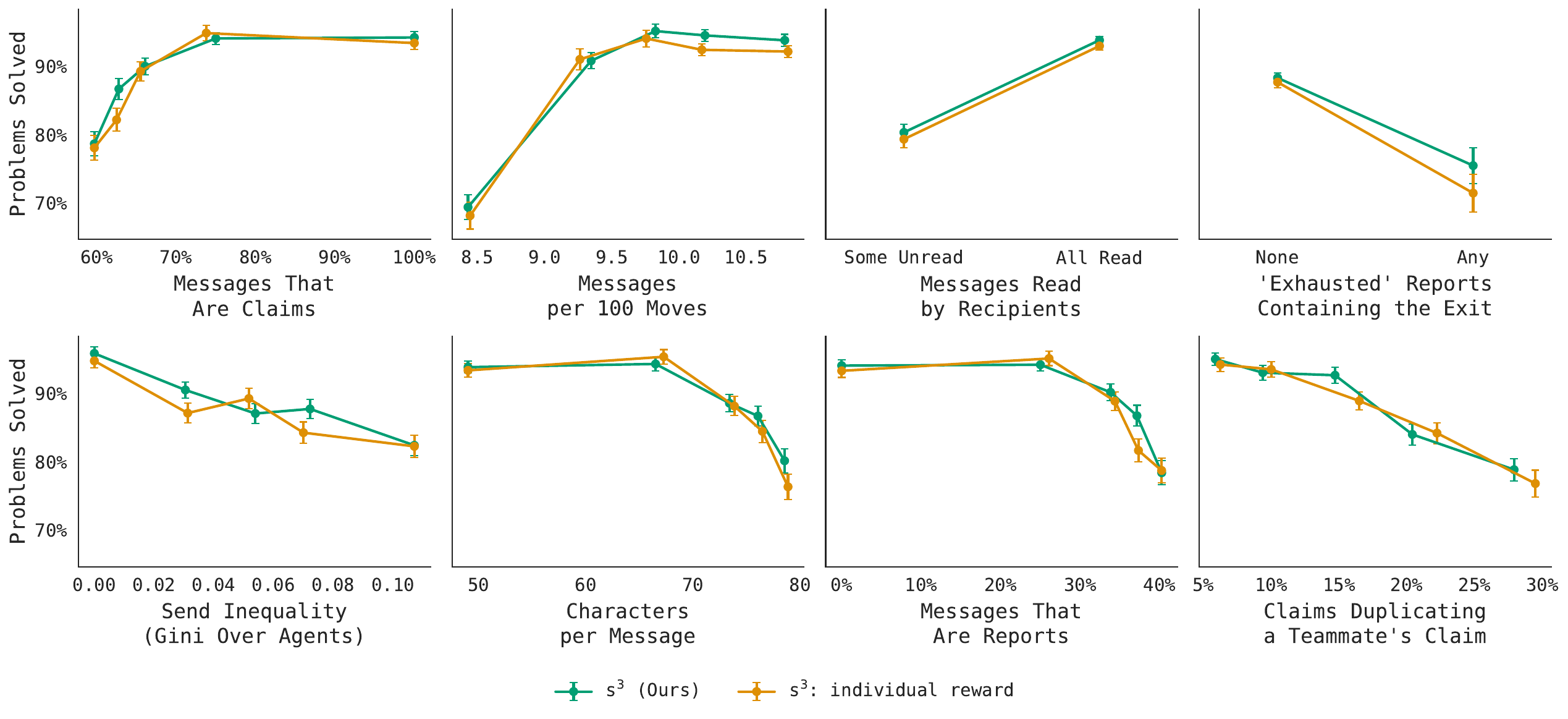}
\caption{Decentralized exploration success by team feature, updates 101--200. Teams are split into fifths of each feature (points at each fifth's median) or, for the two skewed features, into two groups. Success is compared within the same search tree and 10-update bin: each team's success minus the mean of its search tree and bin, plus the overall mean. Panels run from the most positive to the most negative slope under \method{}. Mean $\pm$ SE across teams.}
\label{fig:app-maze-assoc}
\end{figure}

\textbf{What goes with success.} Within the same search tree and 10-update bin (Figure~\ref{fig:app-maze-assoc}), duplicate claims go with $-10.4$ and $-10.9$ points per SD under \method{} and individual reward, unequal sending across agents with $-5.4$, and the share of messages read with $+2.9$ and $+7.5$. A higher share of claims and a lower share of reports go with success ($+7.6$ and $+9.0$), partly because reports accumulate as a search exhausts dead ends. For teams trained from the base model, the message rate and the number of agents that send have associations within $\pm1$ point of zero.

\section{Exploration demonstrations}
\label{app:maze-sft}

\textbf{Scripted explorers.} The demonstrations are generated by a script, not by a model. Four scripted explorers act in a random, non-synchronized order on the real exploration environment, and messages pass through a simulated mailbox with the same format as training. Each explorer runs depth-first search over its own observations. Its state is built only from its own observations and the text of messages it has read; it never sees the search tree layout. At a junction, it announces the branch it will take and prefers branches no teammate has claimed:
\begin{quote}\small
``I'm checking the branch from R1203 through R4410.''
\end{quote}
After every door below a branch has been checked, it reports the branch, stating whether the conclusion rests on its own search alone or also on teammates' reports:
\begin{quote}\small
``Branch from R1203 through R4410 is exhausted; no exit. I followed it and checked every onward door before backtracking.''
\end{quote}
It reads whenever it has unread messages, skips branches teammates have reported as exhausted, and backs out of a branch if a teammate reports it exhausted while it is inside, explaining why in a short reasoning line (``agent\_9 reported R1203 through R4410 exhausted. I am inside that branch; backtrack toward its entry.''). Every message is a broadcast. The independent explorers used for the solo learner's demonstrations run the same search with no messages.

\textbf{Data.} We generate demonstrations on depth-2 training search trees and keep an episode only if the team reached the exit, every agent's supervised tokens fit within the 8,193-token allowance, and every history fits in the 32,768-token context. We keep 512 training and 16 validation episodes. Each agent's private history becomes one training example rendered in the Qwen3 chat format with thinking on, and only the agent's own actions and reasoning lines are supervised. The solo data use the same 512 search trees and the same random seeds, with four independent explorers per search tree.

\textbf{Fine-tuning.} Each fine-tuning stage runs for 50 updates, each on 8 sequences of up to 32,768 tokens packed from these histories, with AdamW, learning rate $10^{-5}$, and no weight decay. The goal is valid tool use and exploration while retaining room for improvement on the larger RL search trees.

\section{Evaluation and metrics}
\label{app:eval}

\textbf{Protocol.} Starting policies and final policies after 200 updates are evaluated on the 32 held-out evaluation problems at each $N\in\{1,2,4,6,8,12,16\}$. Starting policies include the base model and the corresponding SFT policy for conditions fine-tuned on demonstrations. Each problem has one independent attempt for each of six evaluation seeds. Evaluation seeds are distinct from training seeds, and names are resampled in every episode. Each agent gets the training allowance and context. Sampling temperature is 0.7 for decentralized exploration and 1.0 for information aggregation. Communicating policies run as teams. Solo-trained, initial-solo, SFT-solo, and no-communication policies run without messaging, and the population succeeds if any agent succeeds. Each inference server is bound to the exact checkpoint and can serve multiple evaluation sizes; native histories, messages, reads, and failure records are saved.

\textbf{Statistics.} For each seed, we compute solve rate and mean population tokens over the 32 problems, counting task failures. We report the mean over the six seed means and its standard error.

\textbf{Token efficiency.} For each seed, $\mathrm{eff} = \mathrm{SR}\cdot NL/\bar T$, where SR is the solve rate, $NL$ is the team's total allowance, and $\bar T$ is the mean number of tokens the team generated. A team that solves every problem while spending its whole allowance has efficiency 1, and spending half as much doubles it.

\textbf{Message graphs.} From each episode we build a directed graph with an edge from $i$ to $j$ if $i$ sent at least one message that reached $j$; a broadcast adds edges to every teammate. With $k_j$ the in-degree of agent $j$ and $k^{*}=\max_j k_j$, \emph{centralization} is $\sum_j (k^{*}-k_j)/(N-1)^2$, which is 0 when every agent receives from the same number of teammates and 1 when a single agent receives from everyone and no one else receives anything. \emph{Reciprocity} is the fraction of edges whose reverse edge also exists. For decentralized exploration, \emph{room overlap} is $(V-U)/V$, where $V=\sum_i (|\mathcal R_i|-1)$ counts the distinct rooms each agent visited apart from the shared start, and $U=|\bigcup_i \mathcal R_i|-1$ counts the distinct rooms the team visited. It is 0 when no two agents visit the same room and approaches 1 when all agents follow the same path. It includes corridors that several agents must share to reach different branches, so it is not a pure measure of wasted effort.

\end{document}